\documentclass[apjl]{emulateapj}
\usepackage{psfig,amsfonts,amsmath,graphicx,natbib,apjfonts}
\def\ra#1#2#3{#1$^{\rm h}$#2$^{\rm m}$#3$^{\rm s}$}
\def\dec#1#2#3{$#1^\circ#2'#3''$}
\def\swift{{\it Swift}}
\def\nod{\nodata}
\def\trans{NGC\,300\,OT2008-1}
\def\m85{M85\,OT2006-1}

\def\har{1}
\def\hubble{2}
\def\uva{3}
\def\sto{4}
\def\clay{5}
\def\sdsu{6}
\def\dtm{7}
\def\stew{8}
\def\ociw{9}
\def\lund{10}

\begin{document}

\title{An Intermediate Luminosity Transient in NGC\,300: The Eruption
of a Dust-Enshrouded Massive Star}

\author{
E.~Berger\altaffilmark{\har},
A.~M.~Soderberg\altaffilmark{\har,}\altaffilmark{\hubble},
R.~A.~Chevalier\altaffilmark{\uva},
C.~Fransson\altaffilmark{\sto},
R.~J.~Foley\altaffilmark{\har,}\altaffilmark{\clay},
D.~C.~Leonard\altaffilmark{\sdsu},
J.~H.~Debes\altaffilmark{\dtm},
A.~M.~Diamond-Stanic\altaffilmark{\stew},
A.~K.~Dupree\altaffilmark{\har},
I.~I.~Ivans\altaffilmark{\ociw},
J.~Simmerer\altaffilmark{\lund},
I.~B.~Thompson\altaffilmark{\ociw},
C.~A.~Tremonti\altaffilmark{\stew}
}

\altaffiltext{\har}{Harvard-Smithsonian Center for Astrophysics, 60
Garden Street, Cambridge, MA 02138}

\altaffiltext{\hubble}{Hubble Fellow}

\altaffiltext{\uva}{Department of Astronomy, University of Virginia,
P.O. Box 400325, Charlottesville, VA 22904-4325}

\altaffiltext{\sto}{Department of Astronomy, Stockholm University,
AlbaNova, SE-106 91 Stockholm, Sweden}

\altaffiltext{\clay}{Clay Fellow}

\altaffiltext{\sdsu}{Department of Astronomy, San Diego State
University, PA-210, 5500 Campanile Drive, San Diego, CA 92182-1221}

\altaffiltext{\dtm}{Carnegie Institution of Washington, Department of
Terrestrial Magnetism, 5241 Broad Branch Road, Washington, DC 20015}

\altaffiltext{\stew}{Steward Observatory, University of Arizona, 933
North Cherry Avenue, Tucson, AZ 85721}

\altaffiltext{\ociw}{Observatories of the Carnegie Institution
of Washington, 813 Santa Barbara Street, Pasadena, CA 91101}

\altaffiltext{\lund}{ Lund Observatory, Box 43, SE-221 00 Lund,
Sweden}

\begin{abstract} We present multi-epoch high-resolution optical
spectroscopy, UV/radio/X-ray imaging, and archival {\it Hubble} and
{\it Spitzer} observations of an intermediate luminosity optical
transient recently discovered in the nearby galaxy NGC\,300.  We find
that the transient (NGC\,300\,OT2008-1) has a peak absolute magnitude
of $M_{\rm bol}\approx -11.8$ mag, intermediate between novae and
supernovae, and similar to the recent events M85\,OT2006-1 and
SN\,2008S.  Our high-resolution spectra, the first for this event, are
dominated by intermediate velocity ($\sim 200-1000$ km s$^{-1}$)
hydrogen Balmer lines and \ion{Ca}{2} emission and absorption lines
that point to a complex circumstellar environment, reminiscent of the
yellow hypergiant IRC+10420.  In particular, we detect broad
\ion{Ca}{2} H\&K absorption with an asymmetric {\it red} wing
extending to $\sim 10^3$ km s$^{-1}$, indicative of gas infall onto a
massive and relatively compact star (blue supergiant or Wolf-Rayet
star); an extended red supergiant progenitor is unlikely.  The origin
of the inflowing gas may be a previous ejection from the progenitor or
the wind of a massive binary companion.  The low luminosity,
intermediate velocities, and overall similarity to a known eruptive
star indicate that the event did not result in a complete disruption
of the progenitor.  We identify the progenitor in archival {\it
Spitzer} observations, with deep upper limits from {\it Hubble} data.
The spectral energy distribution points to a dust-enshrouded star with
a luminosity of about $6\times 10^4$ L$_\odot$, indicative of a $\sim
10-20$ M$_\odot$ progenitor (or binary system).  This conclusion is in
good agreement with our interpretation of the outburst and
circumstellar properties.  The lack of significant extinction in the
transient spectrum indicates that the dust surrounding the progenitor
was cleared by the outburst.  We thus predict that the progenitor
should be eventually visible with {\it Hubble} if the transient event
marks an evolutionary transition to a dust-free state, or with {\it
Spitzer} if the event marks a cyclical process of dust
formation. \end{abstract}


\section{Introduction}
\label{sec:intro}

In recent years dedicated searches and serendipitous discoveries have
uncovered several optical transients with luminosities intermediate
between the well-studied classes of nova eruptions ($M\sim -8$ mag)
and supernova (SN) explosions ($M\sim -17$ mag).  As can be expected
from the wide gap in luminosity, these intermediate luminosity optical
transients (hereafter, ILOTs\footnotemark\footnotetext{We prefer the
neutral designation {\it intermediate luminosity optical transient}
for this class of events since it does not impose a bias on the
discovery method and initial mis-classification (as in the case of the
``SN impostors'' designation \citealt{vpk+00}), it avoids various
observational cuts on the optical properties (e.g., based on color as
in the case of the ``luminous red nova'' designation \citealt{kor+07})
which may also be due to extrinsic effects (e.g., extinction), it is
less cumbersome and more intuitive than designations based on
prototype events (e.g., V838 Mon-like, $\eta$ Car analogs), and
perhaps most importantly, it makes no assumption about an eruptive
versus explosive origin (as in the case of the ``luminous red novae''
or ``SN impostors'').}) appear to be diverse in their properties and
origins.  Some have been classified as low luminosity core-collapse
SNe, with inferred energies and $^{56}$Ni masses that are at least an
order of magnitude below typical SN events (e.g., \citealt{pzt+04}).
Others, sometimes initially classified as type IIn SNe, have been
subsequently tagged with the catch-all designation of ``SN impostors''
(e.g., \citealt{vpk+00}), that includes events resembling luminous
blue variable (LBV) eruptions.  Finally, a small number of somewhat
dimmer events have been recently grouped under the proposed name
``luminous red novae'' (LRNe; \citealt{kor+07}), based on their red
optical colors; the origin of LRNe and their relation to each other
remain unclear.  Overall, the various groupings have been subject to
debate, and there is no clear agreement about the nature of individual
objects or the degree of overlap between the various designations.

Equally important, the connection between the different types of
events and different classes of progenitors remains unclear.  For
example, some LBV eruptions have been initially classified as SNe
(e.g., SNe 1961V and 1954J; \citealt{gsp+89,vfc+05}).  Similarly, the
progenitor of the recent event SN\,2008S, which was detected in
archival {\it Spitzer Space Telescope} images, has been argued to be
an extreme AGB star \citep{pkt+08,tps+08} or an LBV \citep{sgl+08}.
The nature of the event itself is also unclear, with claims of a low
mass electron-capture SN \citep{tps+08} and an LBV-like outburst
\citep{sgl+08}.  The recent event \m85, and the possibly related
eruptive object V838 Mon, have been speculated to possibly result from
stellar mergers \citep{mhk+02,st06,kor+07}, but other possibilities
have been proposed such as a low-luminosity SN for \m85\
\citep{pds+07}, and an AGB pulse, nova eruption from an embedded
common envelope white dwarf, or planet capture for V838 Mon (e.g.,
\citealt{rzs+06} and references therein).

Regardless of the exact interpretation, it is clear that the various
proposed eruption/explosion scenarios mark important phases in the
evolution of massive stars, possibly on the path to diverse types of
SNe.  Therefore, a mapping between ILOTs and their progenitors will
provide important constraints on the initial conditions (i.e.,
progenitors and circumstellar properties) of SN explosions.  The
present uncertainty in this mapping highlights the need for detailed
observations that can uncover the conditions both prior to and during
the ILOT events.  These include the identification of progenitors, a
detailed study of the circumstellar environment, and measurements of
the event properties across the electromagnetic spectrum.  Such
observations will also allow us to determine the connection between
the various proposed categories of ILOTs, and to map the overall
diversity of this seemingly heterogeneous class.

In this paper we present such a combination of observations for an
ILOT discovered in the nearby galaxy NGC\,300 on 2008 April 24 UT
(hereafter, \trans).  For the first time for such an event, we combine
multi-wavelength observations (UV, optical, radio, and X-rays),
high-resolution echelle spectroscopy, and archival {\it Hubble Space
Telescope} and {\it Spitzer Space Telescope} data to show that \trans\
was the result of an eruption from a $\sim 10-20$ M$_\odot$
dust-obscured and relatively compact star, possibly in a binary
system.  Our high-resolution spectra, the first for this event,
resemble the yellow hypergiant star IRC+10420 (see also
\citealt{bbh+09}), suggesting that \trans\ may mark a transition to a
similar evolutionary stage.  A comparison to the recent events
SN\,2008S and \m85\ shows that all three events may have originated
from the same phenomenon, but with a scatter in the energy release and
circumstellar properties, that may point to a range of progenitor
masses.

\section{Observations}
\label{sec:obs}

\trans\ was discovered by \citet{mon08} on 2008 Apr 24.16 UT in the
outskirts of the nearby galaxy NGC\,300 ($d=1.9$ Mpc;
\citealt{gps+05}) with an apparent brightness of 16.5 mag
(unfiltered).  Subsequent observations on May 14.14 UT revealed that
the object had brightened to about 14.2 mag \citep{mon08}.  \trans\
was not detected in observations from 2008 Apr 17.1 UT, but only to a
limit of 15.5 mag.  It was also not detected in deeper observations
taken on 2007 Dec 30.8 and 2008 Feb 8.75 UT to limiting magnitudes of
18.5 and 18.0 mag, respectively \citep{mon08}.  Since the transient
was brightening at the discovery epoch, we adopt an event start time
that is consistent with the most recent non-detection, $t_0\approx$
Apr 17 UT but note that the actual start time may have occurred
earlier.

A low resolution spectrum of \trans\ obtained by \citet{bwv08} on May
15.4 UT revealed emission lines corresponding to H$\alpha$, H$\beta$,
the \ion{Ca}{2} IR triplet, and the [\ion{Ca}{2}]$\,\lambda\lambda
7293,7325$ doublet (see also \citealt{bbh+09}).  The
marginally-resolved lines indicated velocities of $\lesssim 800$ km
s$^{-1}$, significantly lower than those of both classical novae and
supernovae.

Following the announcement of the discovery and early spectroscopy,
and given the unique opportunity to study a nearby ILOT in detail, we
initiated observations across the electromagnetic spectrum,
particularly high-resolution spectroscopy.  In addition, we obtained
archival {\it Hubble Space Telescope} and {\it Spitzer Space
Telescope} observations at the position of the transient to search for
a pre-discovery counterpart.

\subsection{Astrometry}
\label{sec:astrom}

The original reported position of \trans\ was
$\alpha=$\ra{00}{54}{34.16}, $\delta=$\dec{-37}{38}{28.6} (J2000),
measured relative to the core of NGC\,300 \citep{mon08}.  As part of
our initial spectroscopic observations we obtained an $r$-band image
of the field with the Low Dispersion Survey Spectrograph (LDSS-3)
mounted on the Magellan/Clay 6.5-m telescope.  Using 13 objects in
common with the Naval Observatory Merged Astrometric Dataset (NOMAD),
we measure the actual position of the transient to be
$\alpha=$\ra{00}{54}{34.52}, $\delta=$\dec{-37}{38}{31.6} (J2000) with
an uncertainty of about $0.4''$ in each coordinate.  This position is
$4.3''$ west and $3.0''$ south of the IAUC position.  The revised
coordinates allowed us to search for a progenitor in archival HST
observations, and led \citet{tps+08} to identify a counterpart in
archival {\it Spitzer} observations (see \S\ref{sec:prog}).

\subsection{Optical Spectroscopy}

We obtained multiple low- and high-resolution optical spectra of
\trans\ using LDSS-3 and the Magellan Inamori Kyocera Echelle (MIKE)
spectrograph mounted on the Magellan/Clay 6.5-m telescope.  A log of
the observations, including the spectral coverage and resolution is
provided in Table~\ref{tab:spec}.

The LDSS-3 spectra were reduced using standard IRAF routines, while
rectification and sky subtraction were performed using the method and
software described in \citet{kel03}.  Wavelength calibration was
performed using HeNeAr arc lamps, and air-to-vacuum and heliocentric
corrections were applied.  Flux calibration was performed using
several spectrophotometric standard stars.  The MIKE spectra were
reduced using a custom reduction
pipeline\footnotemark\footnotetext{\tt http://www.ociw.edu/Code/mike}
written in Python.  Wavelength calibration was performed using ThAr
arc lamps, and air-to-vacuum and heliocentric corrections were
applied.

\subsection{Ground-based Optical Imaging}

\begin{figure}
\epsscale{1}
\plotone{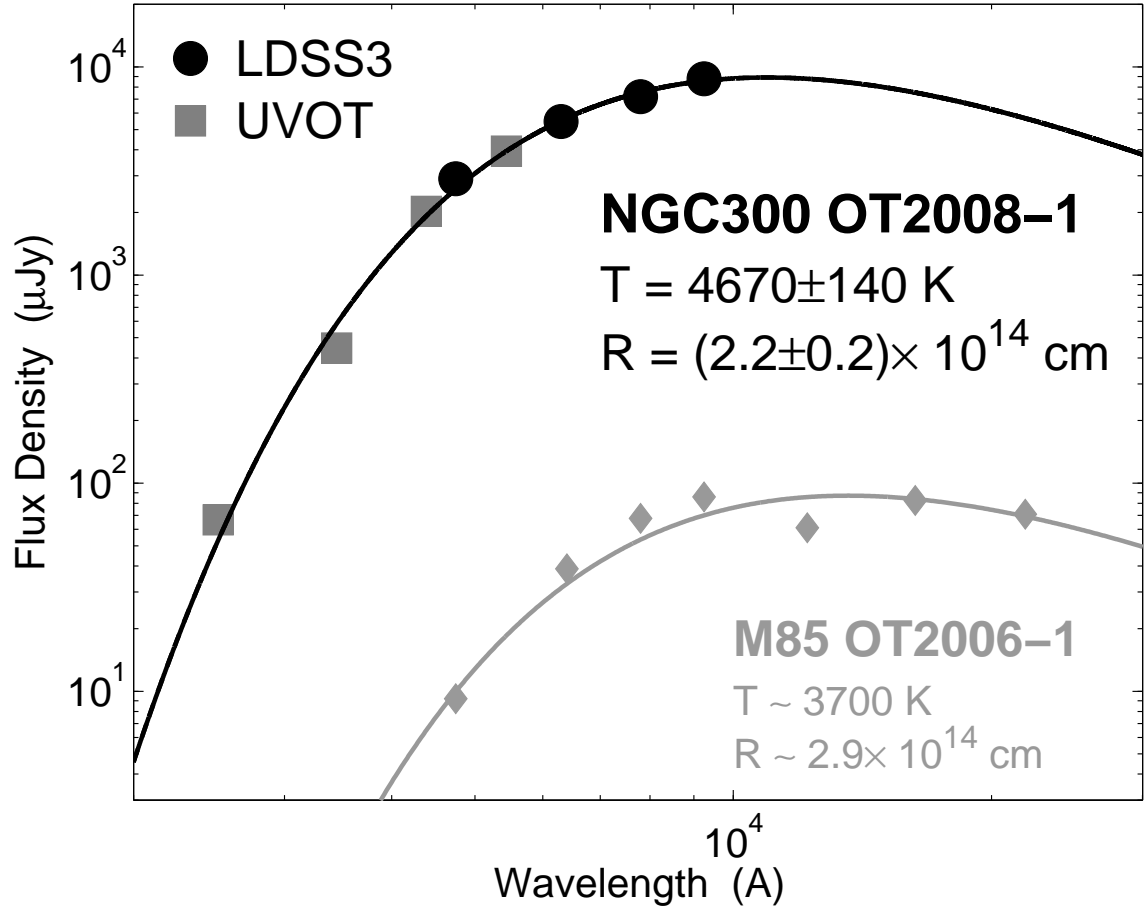}
\caption{Optical (black circles) and UV (gray squares) spectral energy
distribution of \trans\ on May 31.4 UT.  {\it Swift}/UVOT data have
been interpolated to this fiducial epoch using the observed decay
rates (Table~\ref{tab:uvot} and Figure~\ref{fig:lcs}).  The black line
is a blackbody model with the best-fit parameters, $T=4670\pm 140$ K
and $R=(2.24\pm 0.20)\times 10^{14}$ cm.  Using an eruption date of
April 17 UT, we find that the average expansion velocity of the
outflow is about 600 km s$^{-1}$.  Also shown is the SED for \m85\
from \citet{kor+07}, with an approximate blackbody model fit,
indicating properties that closely match those of \trans.
\label{fig:optsed}}
\end{figure}

We obtained optical photometric observations with LDSS-3 in the $griz$
filters on May 31.44 UT; see Table~\ref{tab:opt}.  The data were
reduced using standard routines in IRAF.  Photometry was performed
relative to the standard star field E8-A.  The resulting spectral
energy distribution is shown in Figure~\ref{fig:optsed}.

\subsection{{\it Swift} Ultraviolet and Optical Imaging}

We obtained UV and optical observations with the {\it Swift}
UV/optical telescope spanning from 2008 May 20 to October 15; see
Table~\ref{tab:uvot}.  The data were processed using standard routines
within the HEASOFT package.  Photometry of the transient in the UVW1
and $UBV$ images was performed using the standard $5''$ aperture; the
source was not clearly detected in the UVW2 images.  The light curves
are shown in Figure~\ref{fig:lcs}, and the spectral energy
distribution at the epoch of the LDSS-3 observations is shown in
Figure~\ref{fig:optsed}.

\subsection{X-rays}

Simultaneous with the {\it Swift}/UVOT observations, data were also
collected with the co-aligned X-ray Telescope for a total exposure
time of 20.01 ks.  No source is detected at the position of \trans\ to
a $3\sigma$ limit of $F_X\lesssim 1.2\times 10^{-14}$ erg s$^{-1}$
cm$^{-2}$, where we have assumed a $kT=0.5$ keV thermal bremsstrahlung
model and Galactic absorption, $N_H=3.1\times 10^{20}$ cm$^{-2}$.  A
power law model with a photon index of $-2$ leads to a similar limit
of $\lesssim 1.6\times 10^{-14}$ erg s$^{-1}$ cm$^{-2}$.  The
corresponding limit on the luminosity is $L_X\lesssim 5\times 10^{36}$
erg s$^{-1}$.  This is lower by at least an order of magnitude than
any X-ray detected supernova to date.

\subsection{Radio}

We observed \trans\ with the Very Large
Array\footnotemark\footnotetext{The National Radio Astronomy
Observatory is a facility of the National Science Foundation operated
under cooperative agreement by Associated Universities, Inc.})
beginning on 2008 May 21.65 UT.  No coincident radio source was
detected to a limit of $F_{\nu}\lesssim 90$ $\mu$Jy ($3\sigma$) at
8.46 GHz.  Continued monitoring at frequencies between 4.86 and 22.5
GHz over the subsequent seven months revealed no radio emission, with
the latest non-detection obtained on 2008 November 24 UT
($F_{\nu}\lesssim 110$ $\mu$Jy at 8.46 GHz).  The upper limits imply a
radio luminosity $L_{\rm \nu,rad}\lesssim 3\times 10^{23}$ erg
s$^{-1}$ Hz$^{-1}$.  This limit is at least two orders of magnitude
below the least luminous radio SNe ever observed (type IIp SNe:
\citealt{cfn06}; type Ibc SN\,2002ap \citealt{bkc02}).  It is also two
orders of magnitude below the detected emission from the ``SN
impostor'' SN\,1961V \citep{src+01}.

\section{Basic Properties of \trans}
\label{sec:prop}

The UV/optical spectral energy distribution (SED) on 2008 May 31 is
shown in Figure~\ref{fig:optsed}.  The data have been corrected for
the low Galactic extinction, $E(B-V)=0.013$ mag \citep{sfd98}.  We
find that the SED is well fit by a blackbody spectrum ($\chi^2_r=7.6$
for 5 degrees of freedom) with a temperature of $T=4670\pm 140$ K and
a radius of $R=(2.2\pm 0.2) \times 10^{14}$ cm.  The best-fit value of
the host galaxy extinction is $A_V=0$ mag, with a $1\sigma$ upper
limit of $A_V<0.6$ mag.  As we show below, the low intrinsic dust
extinction is supported by our detection of weak interstellar
\ion{Na}{1} D absorption lines at the redshift of NGC\,300, from which
we infer negligible line of sight extinction, $E(B-V)=0.05\pm 0.05$
mag.

Using an event start date of April 17 and assuming a constant
expansion rate, the inferred velocity is about 600 km s$^{-1}$, in
good agreement with our measured velocity width of the H$\alpha$
emission line, $\delta v\approx 640$ km s$^{-1}$ (see
\S\ref{sec:lowspec}).

The optical/UV ``bolometric'' luminosity of \trans\ on May 31 is
$L_{\rm bol}\approx 1.6\times 10^{40}$ erg s$^{-1}$, or $M_{\rm
bol}\approx -11.8$ mag.  This is about 4 magnitudes brighter than
typical novae, and about $3-10$ mag fainter than SNe (e.g.,
\citealt{kor+07}).  The optical/UV emission fades slowly to $\delta
t\approx 65$ d with a decline rate of about 0.02 mag d$^{-1}$
(Figure~\ref{fig:lcs}).  However, observations at $\delta t\gtrsim
120$ d reveal significant steepening relative to the early decay rate,
indicating a break in the light curves at $\delta t\approx 80-100$ d
(detailed optical and near-IR light curves are presented in
\citealt{bbh+09}).  We return to this point in \S\ref{sec:comp}.

Using a roughly constant luminosity out to $\approx 65$ d, we find
that the total radiated energy is about $10^{47}$ erg.  Therefore, the
outflow kinetic energy is at least as large, indicating an ejected
mass of $\gtrsim 0.03$ M$_\odot$.  The combination of low ejecta
velocity and an intermediate luminosity suggest that \trans\ is not a
true SN explosion.  Indeed, its luminosity is nearly an order of
magnitude lower than even the claimed low-luminosity core-collapse SNe
1994N, 1997D, 1999br, 1999eu, and 2001dc, with $L_{\rm bol}\approx
(5-10)\times 10^{40}$ erg s$^{-1}$ at $t\lesssim 100$ d
\citep{pzt+04}.

\section{Spectroscopic Properties}
\label{sec:lowspec}

The low-resolution spectra of \trans\ are shown in
Figure~\ref{fig:ldss}.  Convolution of the spectrum from May 31 with
the transmission functions of the $griz$ filters indicates that the
flux calibration across the full spectral range is good to better than
$10\%$.  The most pronounced spectral features are the hydrogen Balmer
emission lines, the [\ion{Ca}{2}] doublet and the \ion{Ca}{2} IR
triplet in emission, and \ion{Ca}{2} H\&K in absorption.  We
additionally detect weak \ion{He}{1} and \ion{Fe}{2} emission lines,
as well as \ion{O}{1} and \ion{Na}{1} in absorption.  A full list of
the lines detected at each epoch is provided in Table~\ref{tab:ldss}.

\subsection{Hydrogen Balmer Lines}

The H$\alpha$ line is best fit by a Lorentzian profile with a velocity
width (FWHM) of $\delta v_{\rm FWHM}\approx 640$ km s$^{-1}$; see
Figure~\ref{fig:redlines}.  The full width at continuum intensity
(FWCI) is $\delta v_{\rm FWCI} \approx 3000$ km s$^{-1}$.  The width
of the line is indicates an expansion velocity of about 600 km
s$^{-1}$, in good agreement with our size and eruption date estimates
(\S\ref{sec:prop}).

The H$\alpha$ flux remains relatively unchanged for the first 2
months, with $F\approx 1.1\times 10^{-13}$ erg s$^{-1}$ cm$^{-2}$.
The H$\beta$ line flux is $1.9\times 10^{-14}$ erg s$^{-1}$ cm$^{-2}$,
leading to $F_{\rm H\alpha}/F_{\rm H\beta}\approx 6$.  This value is
well in excess of $F_{\rm H\alpha}/F_{\rm H\beta}\approx 3$ expected
for case B recombination (at $T=5000$ K appropriate for the SED
temperature of \trans; \S\ref{sec:prop}).  A large Balmer decrement
may be due to dust extinction, but as we already showed above (and see
also \S\ref{sec:ext}) the line of sight extinction to \trans\ is
negligible.  Instead, we conclude that the large Balmer decrement is
due to interaction with a high density circumstellar environment,
through a combination of Balmer self absorption and collisional
excitation \citep{du80}.  The observed ratio is indeed in good
agreement with circumstellar interaction models (e.g.,
\citealt{cf94}).

Using the calculations of \citet{du80} for hydrogen emission at high
electron densities, we find that the observed $F_{\rm H\alpha}/ F_{\rm
H\beta}$ ratio requires a density of $n_e\sim 10^{10}-10^{12}$
cm$^{-3}$.  At this range of densities we expect the H$\gamma$ line to
be suppressed to a minimum ratio of $F_{\rm H\gamma}/F_{\rm
H\beta}\approx 0.2$ (compared to 0.45 for case B recombination).  Our
non-detection of H$\gamma$ on May 31, with $F_{\rm H\gamma}/F_{\rm
H\beta}\lesssim 0.2$, and the detection on June 22 with $F_{\rm
H\gamma}/F_{\rm H\beta}\approx 0.3$ support this conclusion.  We note
that the observed ratio $F_{\rm H\delta}/F_{\rm H\beta}\approx 0.3$
(June 22) is somewhat higher than the expected value of about 0.1.  We
thus conclude that the Balmer lines arise from circumstellar
interaction between the event ejecta and the pre-existing environment.

The H$\alpha$ luminosity resulting from circumstellar interaction is
given by $L_{\rm H\alpha}\approx 1.1\times 10^{38}\,n_{e,10}^2\,
R_{14}^3$ erg s$^{-1}$, where we have assumed a shell with $\delta
R/R=0.1$ and selected fiducial parameters for \trans\ as inferred
above ($n_e=10^{10}n_{e,10}$ cm$^{-3}$ and $R=10^{14}R_{14}$ cm).
Comparing with the observed H$\alpha$ luminosity of $4\times 10^{37}$
erg s$^{-1}$, we find that $n_e\sim 10^{10}$ cm$^{-3}$ and $R\sim
10^{14}$, in reasonable agreement with the values inferred from the
Balmer line ratios and the blackbody fit to the SED.  The shocked
circumstellar material has an inferred mass of $\sim 10^{-5}$
M$_\odot$.  For a uniform wind with a constant velocity, $v_w$, this
corresponds to a mass loss rate of $\dot{M}\sim 10^{-3}v_{w,3}$
M$_\odot$ yr$^{-1}$, where $v_w=10^3v_{w,3}$ km s$^{-1}$.  As a sanity
check, we find that the inferred Thomson optical depth for these
parameters is low, $\tau_T\sim 0.1$.

To summarize, the hydrogen Balmer emission line properties, coupled
with the optical/UV spectral energy distribution, are indicative of an
interaction between a low mass shell traveling at about 600 km
s$^{-1}$ with a high density circumstellar medium, resulting from a
progenitor mass loss rate of $\sim 10^{-3}$ M$_\odot$ yr$^{-1}$.

\subsection{\ion{Ca}{2} Lines}
\label{sec:caii}

We next turn to an investigation of the unusual \ion{Ca}{2} permitted
and forbidden lines, which are not generally detected in SN spectra
(particularly at early time).  As shown in Figure~\ref{fig:redlines},
the [\ion{Ca}{2}] lines are significantly narrower than the H$\alpha$
line, with $\delta v_{\rm FWHM}\approx 290$ km s$^{-1}$.  Moreover,
the lines are clearly asymmetric, ranging at continuum intensity from
$-200$ to $+550$ km s$^{-1}$.  We do not find any clear evolution in
the line brightness and profile between May 23 and June 22 UT.

The \ion{Ca}{2} IR triplet lines also exhibit a complex structure,
with an overall symmetric profile, marked by strong absorption near
the line center.  The observed brighter blue wing indicates that the
center of the absorption feature is redshifted relative to the
emission center.  We measure this shift to be about 80 km s$^{-1}$.
The overall width of the lines is $\delta v_{\rm FWHM}\approx 560$ km
s$^{-1}$, while the (unresolved) absorption component has $\delta
v_{\rm FWHM}\approx 500$ km s$^{-1}$.

Finally, the \ion{Ca}{2} H\&K absorption lines exhibit an asymmetric
profile with an extended red wing; see Figure~\ref{fig:bluelines}.
The overall velocity range at continuum intensity in the May 31
spectrum is $-450$ to $+900$ km s$^{-1}$.  The lines appear to be
somewhat narrower in the later spectrum from June 22.  The large
velocity width of the lines indicates that they are not interstellar
in origin, but instead arise in the circumstellar environment.
However, the fact that we observe in absorption gas that is both
inflowing (red wing) and outflowing (blue wing), indicates that the
lines do not arise exclusively from an outgoing wind associated with
the eruption; in that case we would expect only the outflowing (blue)
component to be observed in absorption.

The existence of material inflowing along the line of sight to \trans\
at velocities of up to $900$ km s$^{-1}$ can be explained as the
result of: (i) gas infall from a previous eruption; or (ii) as the
signature of wind outflow from a companion star impinging on the
progenitor of \trans.  In the former scenario, the observed infall
velocity, which we expect to be close to the progenitor's escape
velocity, points to a massive and compact star such as a Wolf-Rayet
star or a blue supergiant; the velocity is much larger than would be
expected for a red giant, a red supergiant, or an asymptotic giant
branch (AGB) progenitor as proposed by \citet{tps+08} and
\citet{bbh+09}.  In the latter scenario, the companion would have to
satisfy the same constraints (massive and compact) in order to launch
a wind with $\sim 10^3$ km s$^{-1}$.  However, as we show in
\S\ref{sec:prog}, the total progenitor system mass is in the range of
$\sim 10-20$ M$_\odot$, making it difficult to accommodate an early-B
or O star companion as required from the $\sim 10^3$ km s$^{-1}$ wind
velocity \citep{lsl95}.

\section{High Resolution Spectroscopy}

The general spectral properties outlined in the previous section are
borne out in much greater detail and complexity in our high resolution
spectra; see Figures~\ref{fig:mike1}--\ref{fig:mike6}.  These are the
first such observations available for \trans.  We further stress that
neither \m85\ nor SN\,2008S, which appear to share similar properties,
have been observed at high spectral resolution so this is the first
chance to investigate the properties of an ILOT in detail.

\subsection{Extinction from the \ion{Na}{1} D Lines}
\label{sec:ext}

Before we proceed with an analysis of the broad lines associated with
the event itself, we use the \ion{Na}{1} D absorption lines to
determine the line of sight extinction within NGC\,300.  In the May 20
spectrum we detect unresolved lines with the expected $2\!:\!1$ ratio
of equivalent widths (0.17 and 0.08 \AA), indicating that they are
optically thin; see Figure~\ref{fig:hei}.  We find no evidence for
variation in the line profiles between the various epochs.  From the
narrowness and lack of variability we conclude that the lines are
interstellar in origin.  The measured equivalent widths correspond to
a negligible line of sight extinction of $E(B-V)=0.05\pm 0.05$ mag
\citep{mz97}.  This inference is in good agreement with the lack of
obvious extinction evidences in the optical/UV spectral energy
distribution.

In the spectra between June 6 and July 2 we further detect broad
\ion{Na}{1} D absorption, with $\delta v_{\rm FWHM}\approx 300$ km
s$^{-1}$ (Figure~\ref{fig:hei}).  The lines are observed in emission
in the spectrum from August 23 with a similar velocity width.  The
maximum absorption equivalent width is about 0.5 \AA\ in the July 2
spectrum, close to the saturation limit of the relation between
extinction and \ion{Na}{1} D equivalent width.  The allowed range of
$E(B-V)$ for this value is $\approx 0.2-0.5$ mag, in rough agreement
with the upper limit of $A_V<0.6$ mag inferred from the SED of \trans\
(\S\ref{sec:prop}).  We stress, however, that the relation between
extinction and \ion{Na}{1} D equivalent width is calibrated in the
interstellar medium, and there is therefore no reason that it should
equally hold in the complex circumstellar environment of \trans,
particularly in the presence of bright UV emission from the eruption
itself.  Indeed, as we show below, we find definitive evidence for
dust destruction, which would undoubtedly alter the nature of the
relation.

\subsection{\ion{Ca}{2} H\&K Lines}
\label{sec:hk}

The broad and asymmetric \ion{Ca}{2} H\&K absorption lines are perhaps
the most striking aspect of the spectrum (Figure~\ref{fig:cahk}).
From the high resolution spectrum obtained on June 6 we find that the
red wing of the lines is best modeled with a Lorentzian profile
centered at $+195$ km s$^{-1}$ relative to the narrow component (the
systemic velocity), and with a width of $\delta v_{\rm FWHM}\approx
700$ km s$^{-1}$; a Gaussian profile with $\delta v_{\rm FWHM}\approx
860$ km s$^{-1}$ provides a reasonable fit to the red wing as well.
An extension of these profiles to the blue wing of the H\&K lines
significantly over-estimates the line absorption
(Figure~\ref{fig:cahk}).  The blue wing itself is well fit by a
Lorentzian with a width of $\delta v_{\rm FWHM}=200$ km s$^{-1}$.

It is unclear whether the two \ion{Ca}{2} H\&K absorption components
(red and blue wings) arise in physically distinct regions (with the
red side that corresponds to the blue component perhaps contributing
to the red wing), or whether they arise from a kinematically
asymmetric distribution of a single absorbing component.  In the
former case, we interpret the line profiles to arise from the presence
of both outflowing and inflowing gas in the circumstellar environment
due to distinct mass loss episodes, fragmentation of a previously
ejected shell, or the prsenence of a binary companion wind.  This
scenario is similar to the model proposed by \citet{hds02} for the
yellow supergiant star IRC+10420, in which recombined gas no longer
responds to radiation pressure, leading to infall at roughly the
escape velocity.  The measured infall velocity of $\sim 10^3$ km
s$^{-1}$ is significantly higher than $\sim 10^2$ km s$^{-1}$ for
absorption features in IRC+10420 \citet{hds02}, indicating that the
progenitor has a higher escape velocity (i.e., it is significantly
more compact).

Subsequent echelle spectra reveal significant evolution in the
\ion{Ca}{2} H\$K line profiles.  First, the red absorption wing
becomes progressively narrower with time, from about $700$ km s$^{-1}$
on June 6 to about $270$ km s$^{-1}$ on July 14.  At the same time,
the blue absorption wing becomes narrower and eventually starts to be
overtaken by emission with a FWCI width of about 500 km s$^{-1}$ so
that the overall line profile eventually resembles a reversed P Cygni
profile.  Both aspects of the line evolution can be explained as the
result of recombination of Ca$^{++}$ gas produced through shocking of
the initially absorbing gas by the eruption's outflowing material.  In
this scenario, the blue-shifted (outflowing) gas and the highest
velocity inflowing gas are located physically closer to the
progenitor, while the lower velocity infalling material will be
eventually overtaken by the shock at a later time.  A hint of this
effect is seen in the low signal-to-noise spectrum from August 23, in
which very little absorption in the red wing of the H\&K lines is
observed.  The same process can also explain the shift from absorption
to emission in the \ion{Na}{1} D lines.

\subsection{\ion{Ca}{2} IR Triplet}

The low-resolution spectra of \trans\ showed that the \ion{Ca}{2}
triplet emission lines are marked by broad absorption features offset
to the red from the emission line center.  At high resolution, the
line profile is more complex, and the absorption is separated into
multiple components; see Figure~\ref{fig:cair}.  In particular, in the
June 6 spectrum there appear to be three distinct absorption
components in addition to a narrow (interstellar) component, with
velocity widths of 200, 100, and 40 km s$^{-1}$.  All three components
are centered redward of the emission line peak, by 30, 200, and 80 km
s$^{-1}$, respectively.  However, in our final spectrum from August
23, the line profiles become simpler, with just a single absorption
component redshifted relative to the emission line center.  This may
again point to the presence of an initially complex circumstellar
environment, which is partially overtaken and shocked by the
outflowing ejecta.

The emission lines are symmetric and have a relatively constant width
of $\delta v_{\rm FWHM}\approx 400$ km s$^{-1}$ in the various epochs.
Since the ion{Ca}{2} IR lines are produced through radiative
de-excitation from an excited state that is populated through
absorption in the \ion{Ca}{2} H\&K lines, we expect the overall
kinematic profiles of the H\&K lines and the IR lines to match.
However, the situation appears to be more complex here since the broad
red absorption wing of the H\$K lines does not appear to have a clear
emission counterpart in the IR lines.

\subsection{[\ion{Ca}{2}] Lines}

Another striking aspect of the echelle spectra is the strongly
asymmetric profile of the [\ion{Ca}{2}] lines, in which the blue wing
is completely missing; see Figures~\ref{fig:caii_1} and
\ref{fig:caii_2}.  The red wing is best fit by a Lorentzian profile
with a width of $\delta v_{\rm FWHM}\approx 170$ km s$^{-1}$ in the
May 20 and June 6 spectra, decreasing to $\delta v_{\rm FWHM}\approx
140$ km s$^{-1}$ in the June 14 and July 2 spectra, and finally
$\delta v_{\rm FWHM}\approx 75$ km s$^{-1}$ in the August 23 spectra.
These lines are therefore significantly narrower than the \ion{Ca}{2}
IR lines and exhibit a more pronounced decrease in width with time.
This, along with the missing blue wing, indicate that the forbidden
lines are produced in a physically distinct region from the IR lines.
Indeed, to avoid collisional excitation and de-excitation out of the
meta-stable excited state that is responsible for the [\ion{Ca}{2}]
emission, the lines need to be produced in a lower density environment
compared to the IR lines and the hydrogen Balmer lines.

While the blue portion of the line is missing, we find no evidence for
a P Cygni profile.  Thus, the flux deficit is not due to absorption of
the line emission by an optically thick outflow as generally observed
for SNe.  Instead, the line asymmetry may reflect an asymmetric
distribution of the low density gas.  In the context of the complex
circumstellar environment inferred from the \ion{Ca}{2} H\&K and IR
lines, it is likely that the red wing is produced in the low density
outer layers of the inflowing gas, with the blue wing emission being
absorbed by intervening material.  This scenario also explains the
lower velocity width compared to the H\&K lines.

The spectra from July 2 and August 23 reveal the emergence of low
level emission on the blue side of the forbidden lines.  This increase
coincides with the emergence of \ion{Ca}{2} H\&K emission on the blue
side of the lines (\S\ref{sec:hk}), and it is likely due to the same
process of Ca$^{++}$ recombination.

Finally, as discussed in \citet{cf94}, we note that the energy
required to ionize Ca$^{+}$ to Ca$^{++}$ from the meta-stable excited
state (which if left undisturbed would produce the [\ion{Ca}{2}]
doublet through radiative de-excitation) corresponds to 1218.9 \AA, or
about 3.2 \AA\ above the Ly$\alpha$ line.  Thus, when the Ly$\alpha$
line is broader than about 800 km/s, the [\ion{Ca}{2}] doublet
emission can be strongly suppressed by ionization.  For narrow
Ly$\alpha$ emission this effect is inefficient and we expect stronger
[\ion{Ca}{2}] emission.  This is indeed the case for \trans, for which
the typical lines widths are $\lesssim 10^3$ km s$^{-1}$.  As a
result, we expect that events like \trans\ will generally exhibit
strong [\ion{Ca}{2}] emission lines.

\subsection{Hydrogen Balmer lines}

The H$\alpha$ line, which appears to have a symmetric profile at low
resolution, exhibits a more complex combination of emission and
absorption when viewed at high spectral resolution; see
Figure~\ref{fig:ha}.  At all epochs we find a narrow absorption
feature close to the line center, with a width of $\delta v_{\rm
fwhm}\approx 40$ km s$^{-1}$.  The absorption feature is redshifted by
about 30 km s$^{-1}$ relative to the interstellar absorption of the
\ion{Ca}{2} lines.  A second, but weaker, absorption component with a
similar velocity width is detected at a velocity of $-130$ km
s$^{-1}$.

The overall emission profile can be fit with either a Lorentzian
profile or a combination of narrow and broad Gaussian profiles, which
are slightly offset in velocity.  The motivation behind the latter
model is that the broad wings of the line appear to be somewhat
asymmetric, with a slight deficit in the blue portion of the line.  In
the Lorentzian case, the line width remains relatively unchanged
between May 20 and June 14 with $\delta v_{\rm fwhm}\approx 560$ km
s$^{-1}$, and decreases mildly to $\delta v_{\rm fwhm}\approx 440$ km
s$^{-1}$ on July 2 and August 23.  In the Gaussian model the narrow
component appears to increase in width from about 295 km s$^{-1}$ on
May 20 to about 400 km s$^{-1}$ in the spectra from June 6 to August
23.  The broad component remains relatively unchanged with $\delta
v_{\rm fwhm}\approx 1100$ km s$^{-1}$.  With only a mild change in
width, and an increase in brightness relative to the continuum level,
the H$\alpha$ equivalent width increases significantly with time,
ranging from about 10 to 300 \AA\ between May 20 and August 23.

The H$\beta$ line exhibits the same profile as H$\alpha$; see
Figure~\ref{fig:hb}.  In the second epoch (which has a higher signal
to noise ratio), the line is marked by a deep absorption feature with
$\delta v_{\rm fwhm}\approx 70$ km s$^{-1}$, which is centered about
80 km s$^{-1}$ redward of the emission line center.  This component is
thus wider than its counterpart H$\alpha$ absorber, and it also
exhibits a more pronounced shift relative to the line center.  The
H$\beta$ line is again well fit by either a Lorentzian profile or a
combination of narrow and wide Gaussians with offset line centers.  In
the former case, the line width remains unchanged with $\delta v_{\rm
fwhm}=380$ km s$^{-1}$ until July 14, and decreases to about 330 km
s$^{-1}$ on August 23.  In the Gaussian model, the line widths remain
unchanged at $350$ and $1200$ km s$^{-1}$.

Finally, the H$\gamma$ and H$\delta$ lines exhibit similar profiles to
H$\alpha$ and H$\beta$.  In particular, the H$\gamma$ line exhibits
blue-shifted weak absorption similar to that seen in H$\alpha$; see
Figures~\ref{fig:balmer3} and \ref{fig:balmer4}.

\subsection{Additional Absorption and Emission Lines}

In addition to the strong hydrogen Balmer lines and \ion{Ca}{2} lines,
we also detect the \ion{He}{1}$\,\lambda 5875$ emission line
(Figure~\ref{fig:hei}) and the \ion{O}{1} triplets at 7774 and 8446
\AA\ in absorption (Figure~\ref{fig:oi}).  In all cases the line
widths are about 300 km s$^{-1}$.  Moreover, the \ion{O}{1} lines are
clearly asymmetric with broad blue-shifted absorption extending to
about 300 km s$^{-1}$.  This is the opposite effect compared to the
\ion{Ca}{2} H\&K lines, indicating that the absorption in these two
species arises in distinct kinematic components.

In addition, we detect narrow absorption features of \ion{Ca}{2} H\&K,
the \ion{Ca}{2} IR triplet, and \ion{O}{1}, with a typical width of
$\delta v_{\rm FWHM}\approx 25$ km s$^{-1}$.  It is unclear whether
these narrow features are interstellar in origin, or are due to
previous mass loss from the progenitor of \trans, with a much lower
wind velocity (possibly during a preceding red supergiant phase).

\section{The Progenitor of \trans}
\label{sec:prog}

Our precise astrometry of \trans\ (\S\ref{sec:astrom}) enabled a
search for the progenitor in archival {\it Hubble Space Telescope} and
{\it Spitzer Space Telescope} observations (see also
\citealt{tps+08,bbh+09}).

\subsection{Archival {\it Hubble Space Telescope}}
\label{sec:hstdata}

The location of \trans\ was observed with the Advanced Camera for
Surveys (ACS) on 2006 November 8 UT in the F475W (1488 s), F606W (1515
s), and F814W (1542 s) filters as part of program 10915 (PI:
Delcanton).  We retrieved the drizzled images from the HST archive and
performed an astrometric tie relative to our initial LDSS-3 image
(\S\ref{sec:astrom}).  The resulting astrometry has an rms of 30 mas
in each coordinate.

We do not detect any objects at the position of \trans\ to the
following $5\sigma$ limits: $m_{AB}>28.1$ mag (F475W), $m_{AB}>27.8$
mag (F606W), $m_{AB}>27.3$ mag (F814W).  The nearest object to the
position of \trans\ is located about $0.19''$ away, or about 1.75 pc
at the distance of NGC\,300.  Images of the field at the location of
\trans\ are shown in Figure~\ref{fig:hst}.

\subsection{Archival {\it Spitzer Space Telescope}}
\label{sec:spitzerdata}

The location of \trans\ was observed with the Infra-Red Array Camera
(IRAC) on 2007 December 28 UT, and with the Multiband Imaging
Photometer for Spitzer (MIPS) on 2007 July 6 and 16 UT.  We retrieved
the post-BCD data from the {\it Spitzer} archive and performed an
astrometric tie relative to our initial LDSS-3 image
(\S\ref{sec:astrom}).  The resulting astrometry has an rms of $0.3''$
in each coordinate.  Images of the field at the location of \trans\
are shown in Figures~\ref{fig:spitzer1} and \ref{fig:spitzer2}.

As noted by \citet{atel1550} based on our reported astrometry of the
transient \citet{atel1544}, a coincident object is detected in the
{\it Spitzer} data (see also \citealt{tps+08}).  We performed
photometry of this object using a 2-pixel aperture on the 3.6 and 4.5
$\mu$m images, and a 3-pixel aperture on the 5.8, 8.0, and 24 $\mu$m
images.  The aperture size was selected to reduce contamination from
nearby objects.  We further used the standard zeropoints and aperture
corrections provided in the {\it Spitzer} manual.  The resulting
spectral energy distribution of the coincident source, along with the
HST upper limits, is shown in Figure~\ref{fig:spitzer_hst}.

\subsection{Source Properties}

A detailed discussion of the progenitor is presented by \citet{tps+08}
who conclude that it is an extreme AGB star during a short-lived phase
($\sim 10^4$ yr) of its evolution.  Below we provide an independent
analysis of the {\it Spitzer} and HST observations.  The spectral
energy distribution is shown in Figure~\ref{fig:spitzer_hst}.  We find
that the SED is well fit by a blackbody profile ($\chi^2_r\approx 1.5$
for 3 degrees of freedom), indicating that the progenitor was
enshrouded by dust prior to the eruption.  The best-fit blackbody
parameters are $T=340\pm 10$ K and $R=330\pm 20$ AU, with a resulting
luminosity of $5.7\times 10^4$ L$_\odot$.  The total dust mass can be
inferred by setting $\tau=\kappa\,\rho\,R\approx 1$, which for
$\kappa\approx 10$ cm$^2$ g$^{-1}$ \citep{pkt+08}, corresponds to
$5\times 10^{-3}$ M$_\odot$.

The fact that the progenitor was completely obscured by dust, while
\trans\ exhibits negligible dust extinction ($A_V<0.6$ mag;
\S\ref{sec:prop}) demonstrates that the eruption destroyed the
obscuring dust.  Since the size of the dust-obscured region exceeds
the shock radius of \trans\ by about a factor of 20
(\S\ref{sec:prop}), we conclude that the dust was destroyed through
sublimation rather than shock heating.  This process has been
investigated by several authors, and we use here the formulation of
\citet{wd00} developed in the context of gamma-ray bursts.  The dust
destruction radius is given by $R_{\rm sub}\approx 1.2\times
10^{15}L_{40}^{1/2}$ cm, for a typical grain size of 0.1 $\mu$m and a
dust absorption efficiency factor of order unity \citep{wd00}.  For
the radius of the dust-obscured region inferred above, we find that
the luminosity required for sublimation is $L\approx 2\times 10^{41}$
erg s$^{-1}$.  This is about an order of magnitude larger than the
bolometric luminosity measured for \trans\ on May 31 ($\delta t\approx
45$ d).  The measured decay rate of about 0.02 mag d$^{-1}$
extrapolated back to the estimated eruption date is not sufficient to
overcome this discrepancy.  We therefore conclude that either the
early emission from \trans\ was characterized by a higher photospheric
temperature, or the initial event produced a relatively bright UV
flash with a luminosity of about $2\times 10^{41}$ erg s$^{-1}$.

In either case, the destruction of the obscuring dust may be
responsible for the strong \ion{Ca}{2} emission, since the calcium is
initially strongly depleted on dust grains.

\section{Interpretation: The Eruption of a Massive Star}

Our detailed spectroscopic observations of \trans, and the detection
of its progenitor system in archival data, allow us to construct a
basic model of the event and its environment.  The key observational
results presented in the preceding sections are summarized as follows:

\begin{itemize}

\item The line of sight extinction is negligible as inferred from the
UV/optical SED and the weak \ion{Na}{1} D lines.

\item Prior to the event, the progenitor system was enshrouded by
dust, and had a luminosity of about $6\times 10^4$ L$_\odot$,
corresponding to a $\sim 10-20$ M$_\odot$ star.

\item The wide and asymmetric \ion{Ca}{2} H\&K absorption lines
require gas infall at velocities of about $10^3$ km s$^{-1}$, and
hence a massive and relatively compact progenitor or companion.

\item The \ion{Ca}{2} IR lines and the hydrogen Balmer lines are
symmetric and broad ($\sim 300-600$ km s$^{-1}$), and marked by narrow
absorption redward of the emission line peak.

\item The [\ion{Ca}{2}] lines are extremely asymmetric, and
significantly narrower than the other \ion{Ca}{2} lines and hydrogen
Balmer lines.

\item None of the lines exhibit velocities in excess of $\sim 10^3$ km
s$^{-1}$, or P Cygni profiles.

\end{itemize}

These properties bear a striking resemblance to those of the massive
yellow hypergiant IRC+10420 (e.g.,
\citealt{dej98,oud98,hds02,sgl+08,bbh+09}), which is thought to be
rapidly transitioning from a red supergiant phase to a luminous blue
variable or Wolf-Rayet phase.  Like the progenitor of \trans,
IRC+10420 is marked by a bright infrared excess resulting from a dusty
nebula, but unlike \trans\ the central object is not completely
obscured.  The luminosity of IRC+10420, $\approx 5\times 10^5$
L$_\odot$, is about an order of magnitude larger than our inferred
luminosity of the progenitor of \trans\ (\S\ref{sec:prop}).  This
points either to a lower mass or a somewhat earlier evolutionary stage
for the progenitor of \trans.  Equally important, IRC+10420 exhibits
essentially identical hydrogen Balmer and \ion{Ca}{2} IR line profiles
to those seen in our high resolution spectra, including the redshifted
narrow absorption feature (see Figures 4 and 5 of \citealt{hds02} and
Figures 3 and 4 of \citealt{oud98}).

At the same time, there are some distinct difference between the
spectroscopic properties of \trans\ and IRC+10420.  First, the
ion{Ca}{2} H\&K lines in IRC+10420 are symmetric and significantly
narrower than those observed in \trans\ \citep{oud98}.  Second, the
[\ion{Ca}{2}] emission lines in IRC+10420 appear to be symmetric
\citep{oud98,hds02}, although like in the case of \trans\ they are
significantly narrower than the hydrogen Balmer lines and the
\ion{Ca}{2} IR lines.  Third, the \ion{Ca}{2} IR lines, which extend
to about $\pm 100$ km s$^{-1}$ \citep{hds02}, are significantly
narrower than in \trans.  Finally, while the H$\alpha$ profiles in
both objects are generally similar, the red and blue emission peaks of
IRC+10420 have a similar width to the central absorption
\citep{hds02}, whereas in the case of \trans\ the emission peaks are
significantly wider.  In both cases the H$\alpha$ line exhibit a broad
wing, which is interpreted as the effect of electron scattering in the
case of IRC+10420 \citep{hds02}.  In the spectrum of \trans, however,
the core of the line is broader and the wings are less pronounced, and
we interpret the profile as due to circumstellar interaction of an
outflow with a velocity of about 600 km s$^{-1}$ (as opposed to a
outflowing wind velocity of only 50 km s$^{-1}$ inferred for
IRC+10420).

The combination of these various properties suggests that in broad
terms the progenitor of \trans\ may be similar to IRC+10420, but it
may have been in a somewhat different evolutionary stage prior to the
eruption, with still complete dust obscuration.  In addition, the
eruption itself appears to have produced a significantly faster, more
energetic, and more luminous outflow compared to the ``quiescent''
state of IRC+10420.  Thus, it is possible that \trans\ marks the
initial transition to an IRC+10420-like object, and that as the
eruption fades, its properties will converge to those of IRC+10420.

\section{Comparison with \m85\ and SN\,2008S}
\label{sec:comp}

\begin{figure}
\epsscale{1}
\plotone{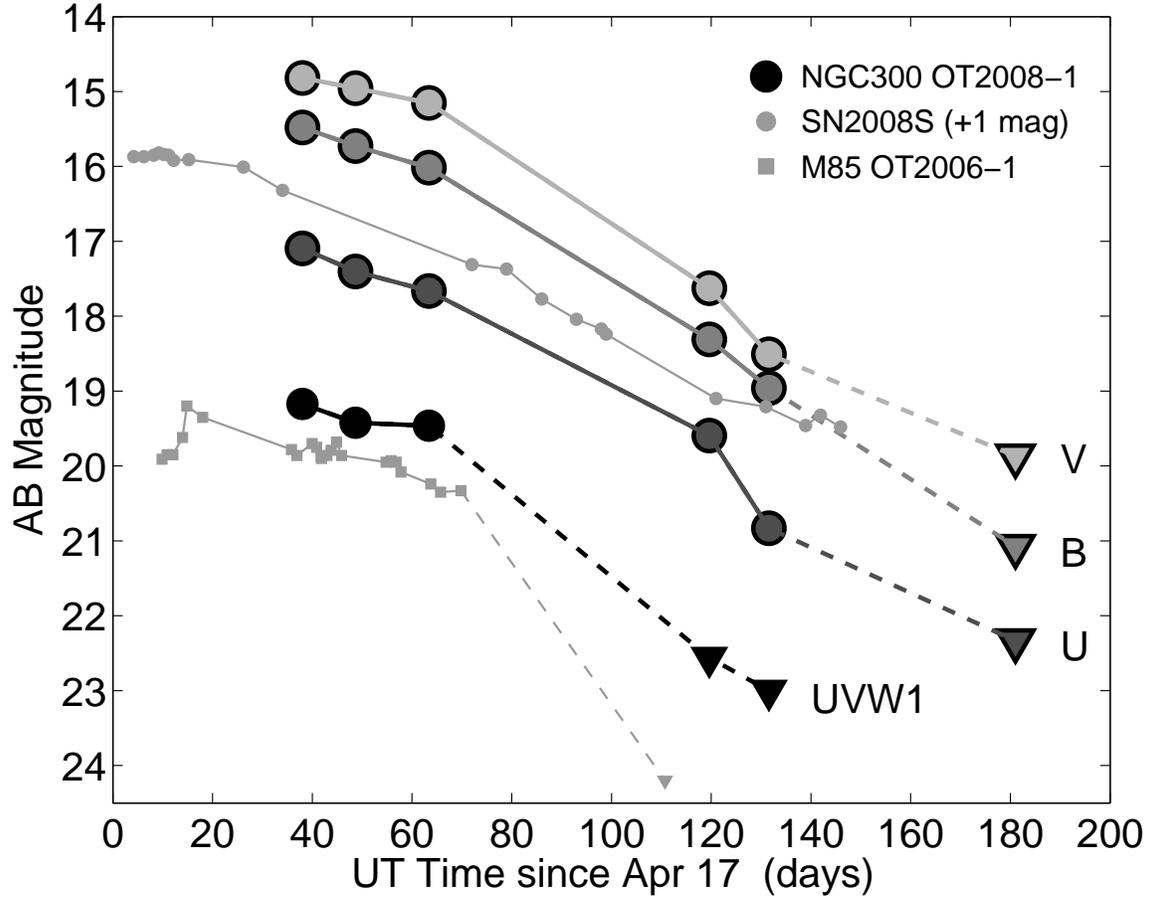}
\caption{\swift/UVOT optical and UV light curves of \trans\ plotted
relative to the fiducial eruption date of April 17 UT.  The initial
decay rate is about 0.02 mag d$^{-1}$ for the first $\sim 65$ d,
followed by significant steepening at about $80-100$ d.  Also shown
are the $R$-band light curves of SN\,2008S (gray circles) and \m85\
(gray squares).  The temporal evolution of all three events is
remarkably similar.
\label{fig:lcs}}
\end{figure}

\begin{figure}
\epsscale{1}
\plotone{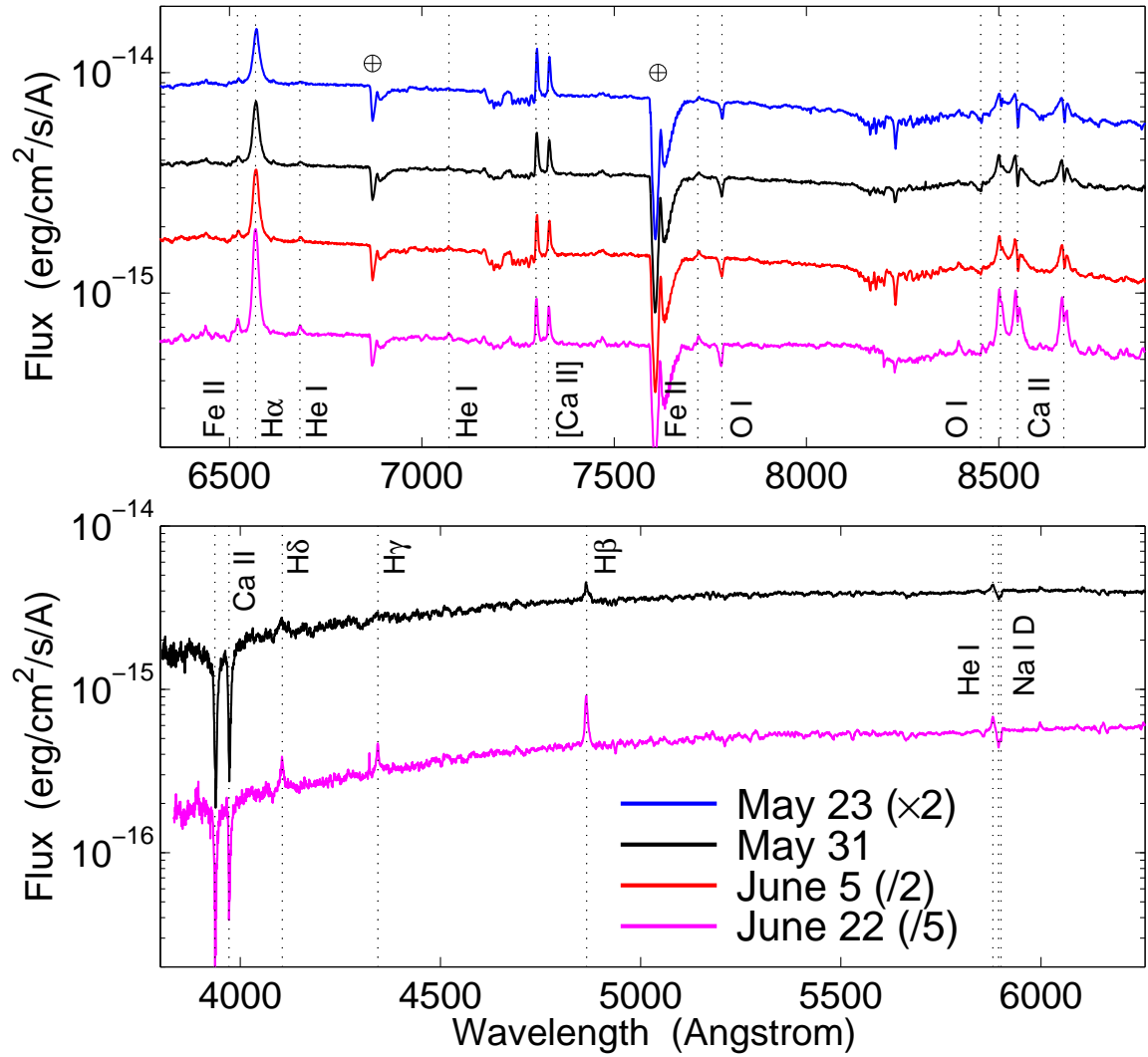}
\caption{Low resolution spectra of \trans\ obtained with the LDSS-3
instrument on the Magellan/Clay 6.5-m telescope.  {\it Top:} Red
spectra with the dominant emission and absorption features identified.
{\it Bottom:} Blue spectra with the dominant emission and absorption
features identified.
\label{fig:ldss}} 
\end{figure}

\begin{figure}
\epsscale{1}
\includegraphics[angle=270,width=7.0in]{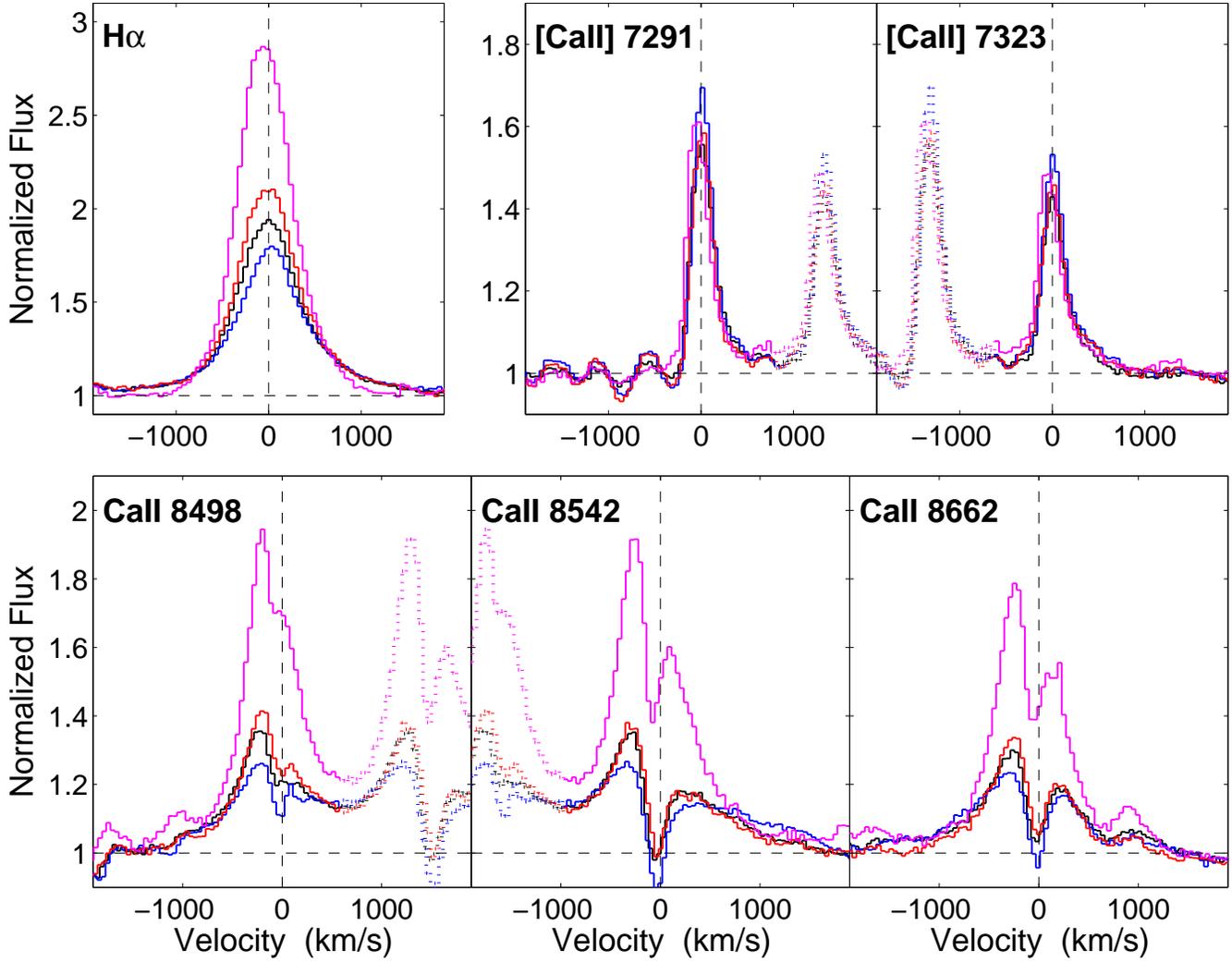}
\caption{Velocity profiles of the dominant emission lines in the red
low-resolution spectra obtained with LDSS-3 (line colors are as in
Figure~\ref{fig:ldss}).  The H$\alpha$ and \ion{Ca}{2} IR triplet
lines exhibit an increase relative to the continuum, while the
[\ion{Ca}{2}] lines remain largely unchanged.  
\label{fig:redlines}}
\end{figure}

\begin{figure}
\epsscale{1}
\includegraphics[angle=270,width=7.0in]{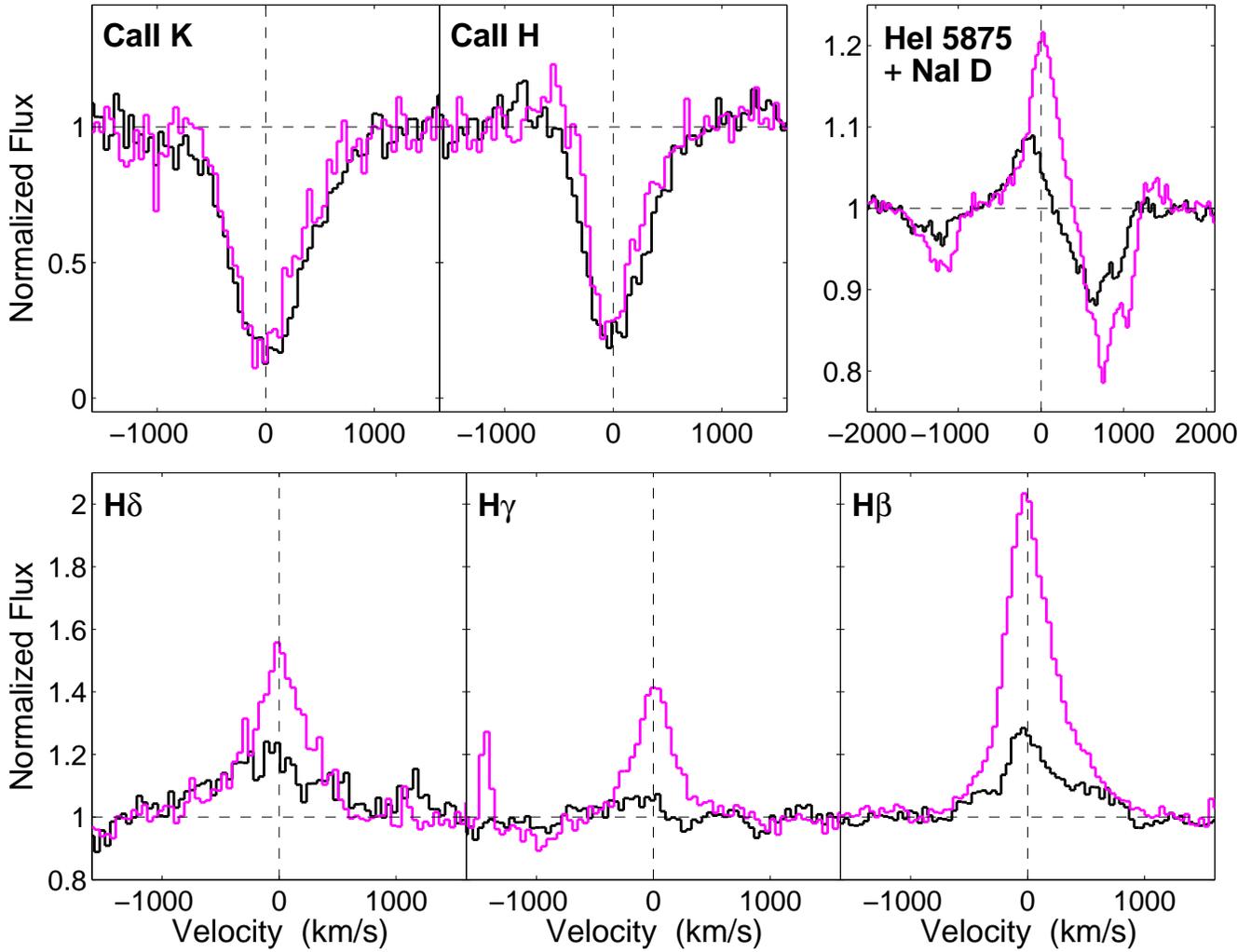}
\caption{Velocity profiles of the dominant emission and absorption
lines in the blue low-resolution spectra obtained with LDSS-3 (line
colors are as in Figure~\ref{fig:ldss}).  The \ion{Ca}{2} H\&K lines
are somewhat narrower in the later spectrum, while the hydrogen Balmer
lines and the \ion{He}{1}$\,\lambda 5877$ line exhibit a significant
increase relative to the continuum brightness.  Similarly, the
\ion{Na}{1} D broad components exhibit increased absorption.
\label{fig:bluelines}}
\end{figure}

\begin{figure}
\epsscale{1}
\includegraphics[angle=90,width=7in]{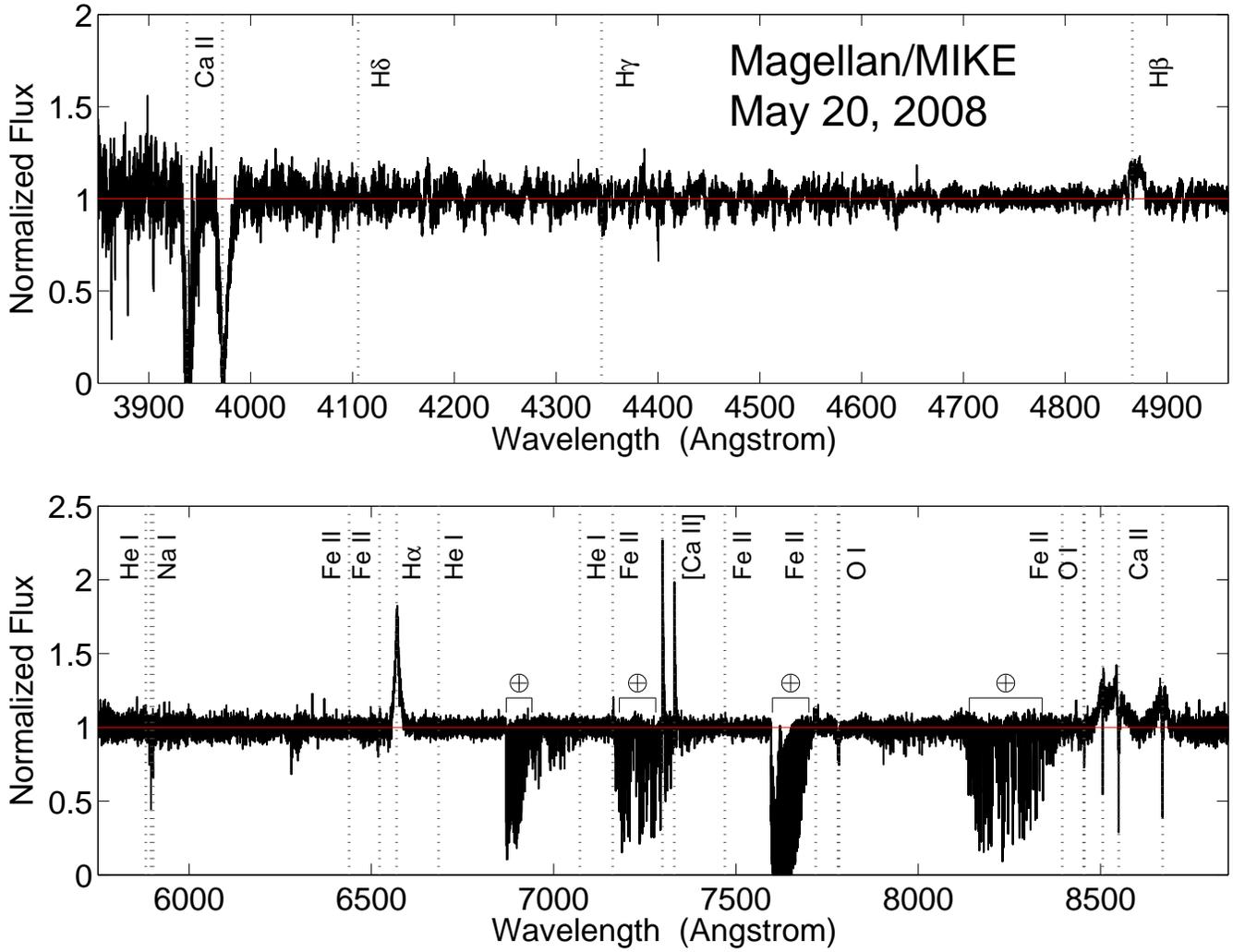}
\caption{High resolution echelle spectrum of \trans\ obtained with the
MIKE spectrograph mounted on the Magellan/Clay 6.5-m telescope on May
20.4 UT.  The prominent lines are marked at a redshift of 200 km
s$^{-1}$.
\label{fig:mike1}}
\end{figure}

\begin{figure}
\epsscale{1}
\includegraphics[angle=90,width=7in]{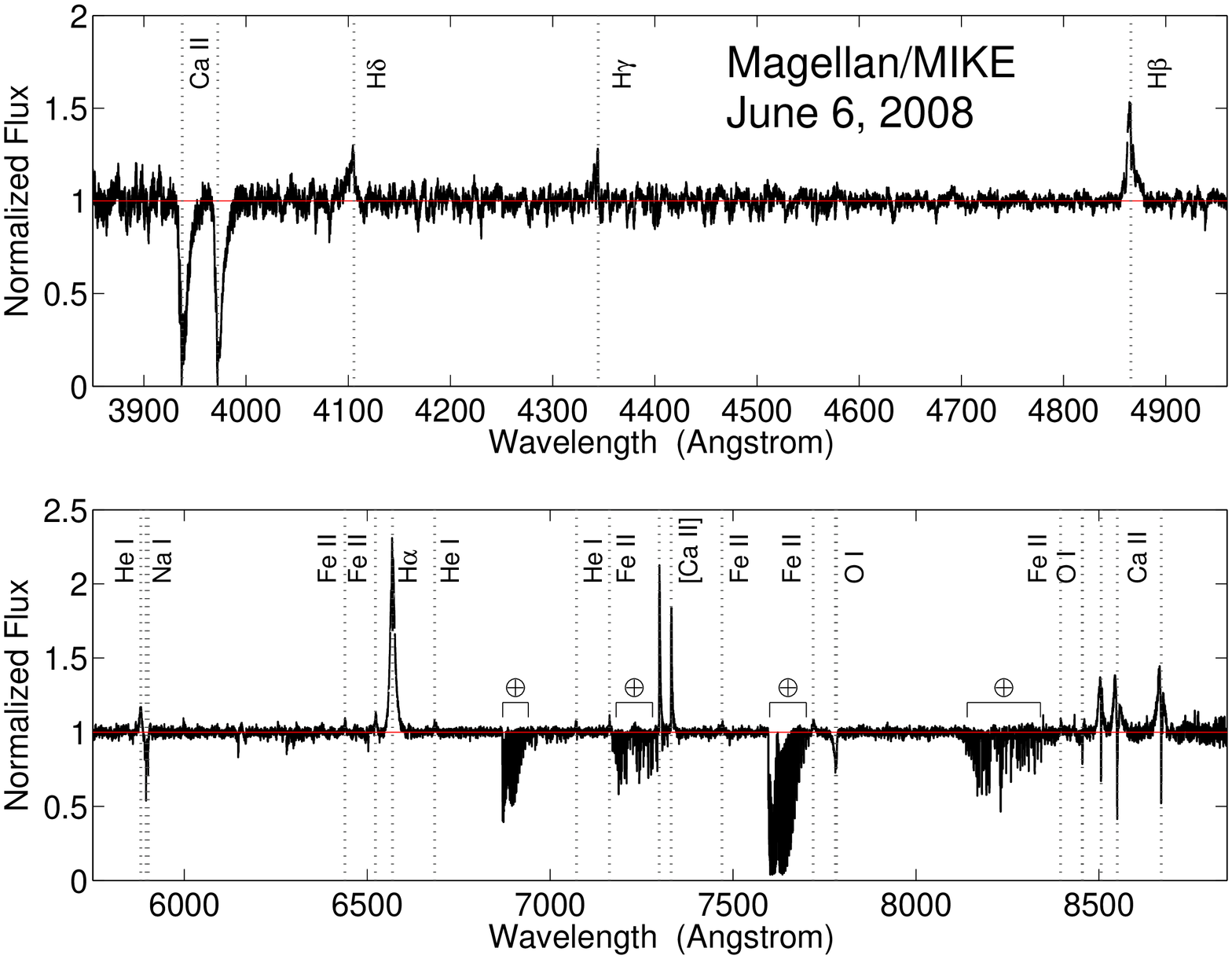}
\caption{Same as Figure~\ref{fig:mike1} but for a spectrum obtained on
June 6.4 UT.  
\label{fig:mike2}}
\end{figure}

\begin{figure}
\epsscale{1}
\includegraphics[angle=90,width=7in]{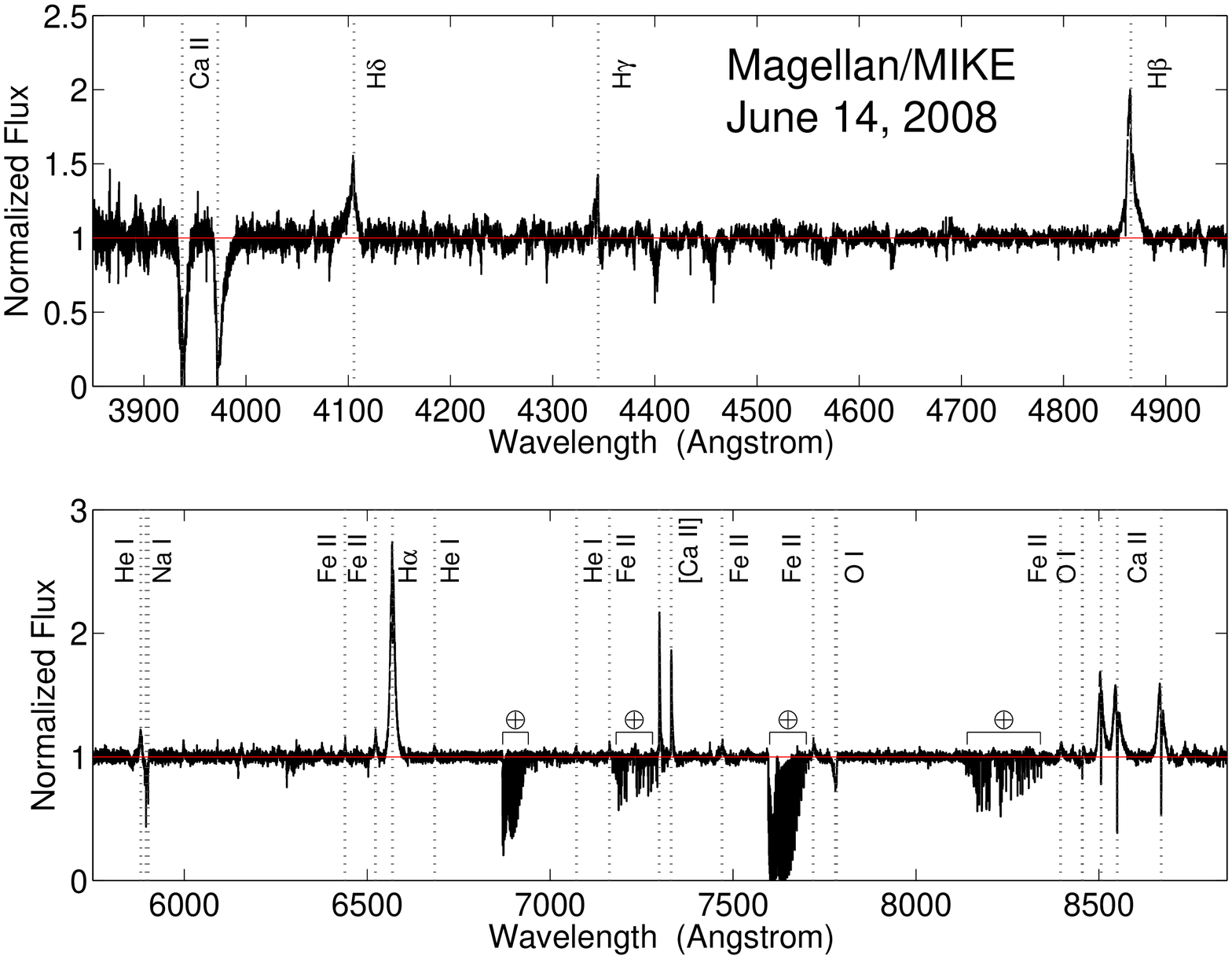}
\caption{Same as Figure~\ref{fig:mike1} but for a spectrum obtained on
June 14.4 UT.  
\label{fig:mike3}}
\end{figure}

\begin{figure}
\epsscale{1}
\includegraphics[angle=90,width=7in]{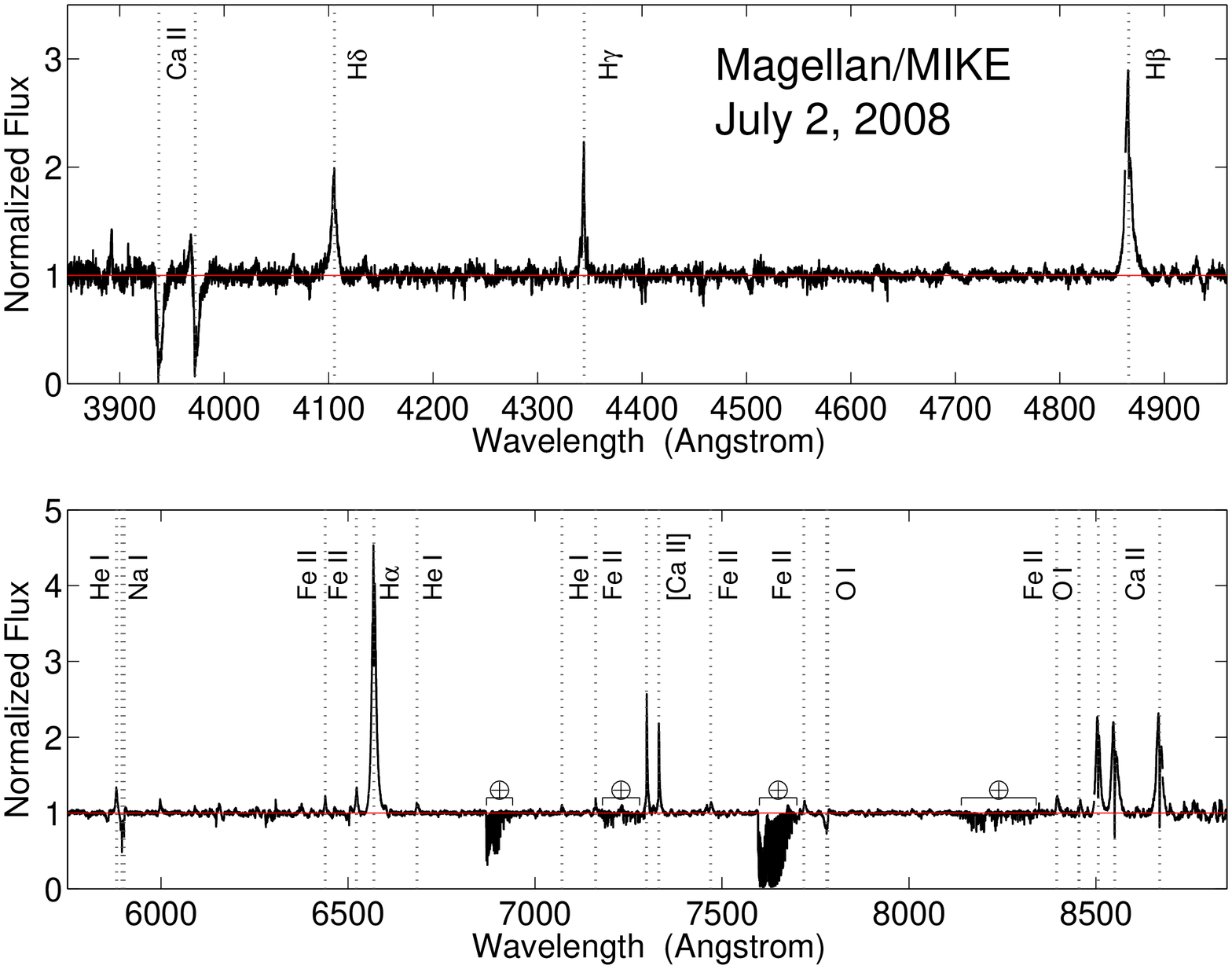}
\caption{Same as Figure~\ref{fig:mike1} but for a spectrum obtained on
July 2.3 UT.  
\label{fig:mike4}}
\end{figure}

\begin{figure}
\epsscale{1}
\includegraphics[angle=90,width=7in]{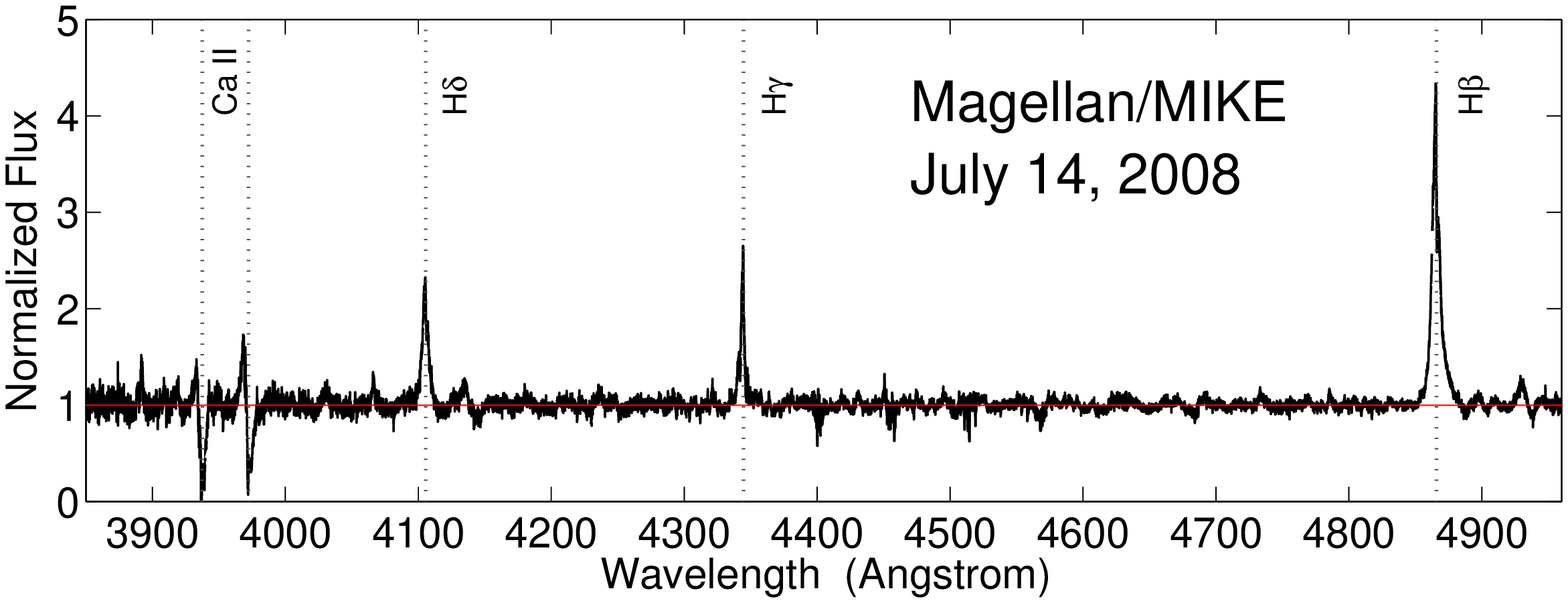}
\caption{Same as Figure~\ref{fig:mike1} but for a spectrum obtained on
July 14.4 UT.  
\label{fig:mike5}}
\end{figure}

\begin{figure}
\epsscale{1}
\includegraphics[angle=90,width=7in]{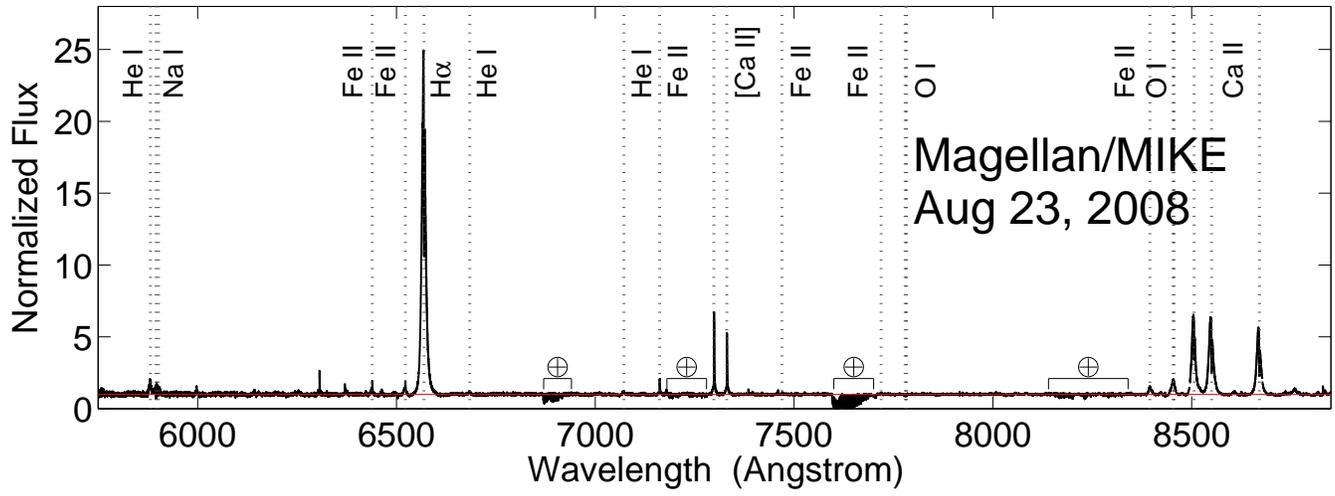}
\caption{Same as Figure~\ref{fig:mike1} but for a spectrum obtained on
August 23.3 UT.  
\label{fig:mike6}}
\end{figure}

\begin{figure}
\epsscale{1}
\plotone{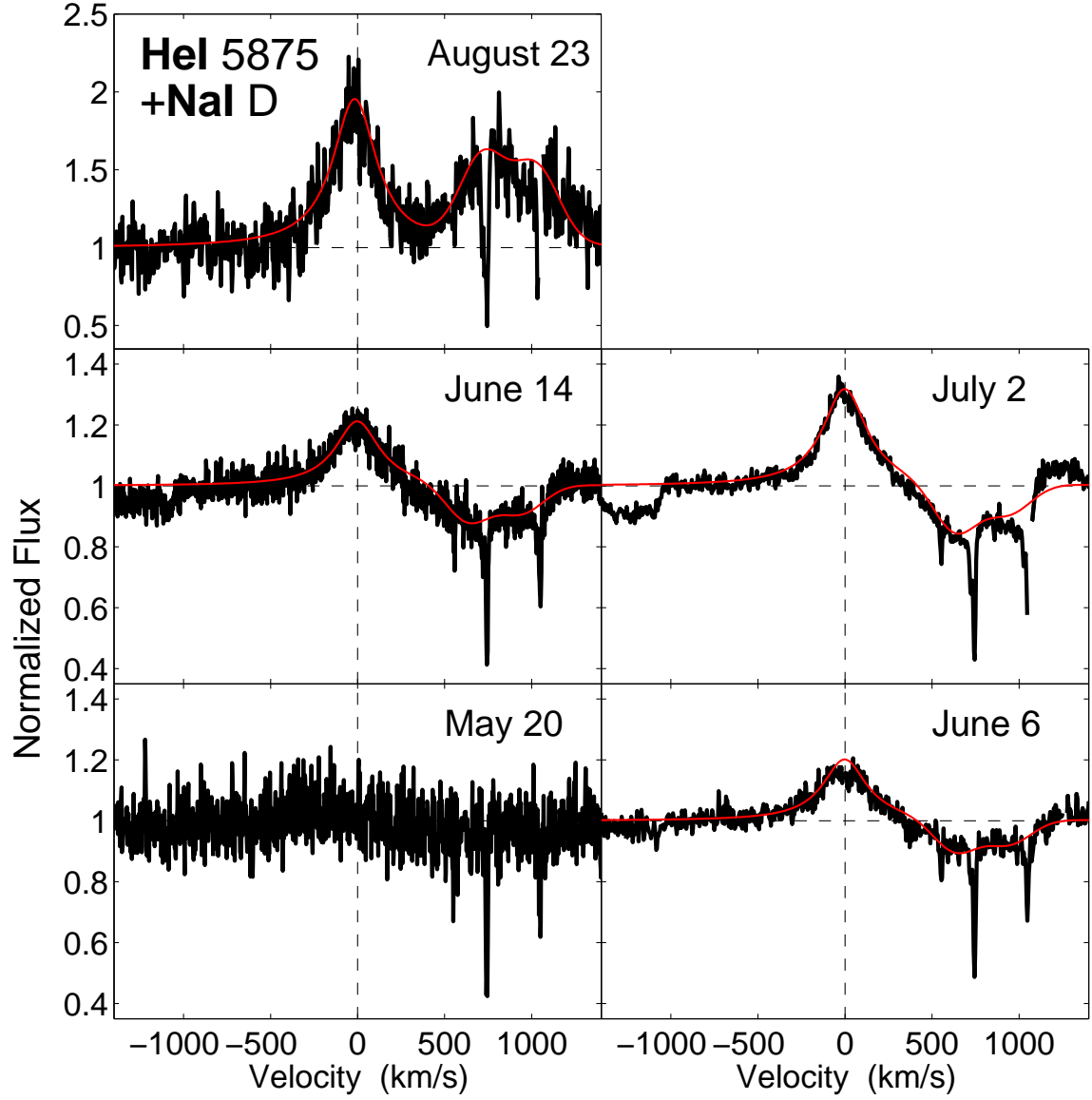}
\caption{\ion{He}{1}$\,\lambda 5875$ and \ion{Na}{1} D line profiles
from the high-resolution spectra.  The \ion{He}{1} emission follows
the same trend of increasing brightness relative to the continuum that
is observed in other emission lines of \trans\ (e.g.,
Figure~\ref{fig:ha}).  The ion{Na}{1} D absorption exhibits both a
narrow interstellar component and a wider component likely associated
with the circumstellar environment.  The broad component changes to
emission in the final spectrum.
\label{fig:hei}}
\end{figure}

\begin{figure}
\epsscale{1}
\plotone{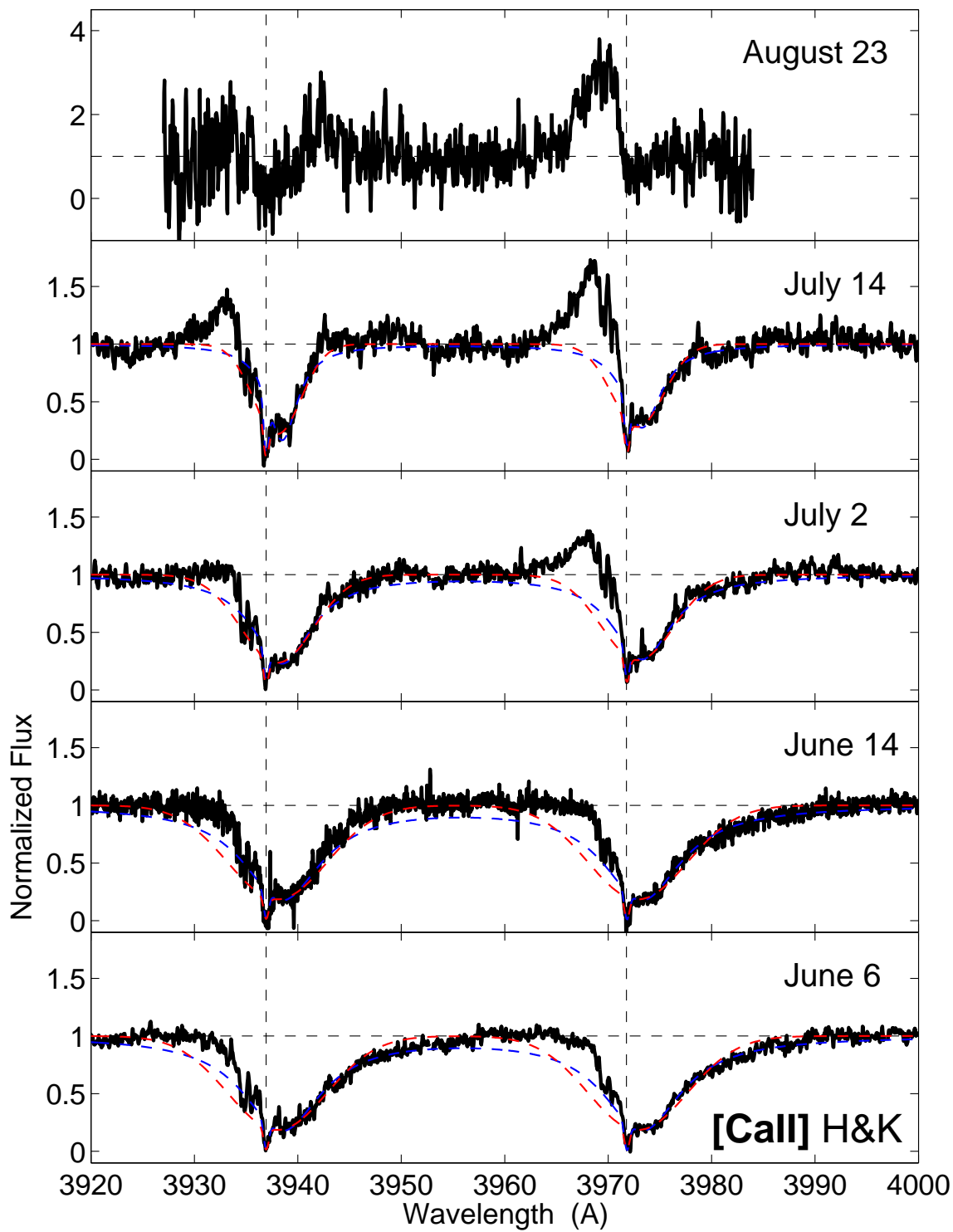}
\caption{Absorption profiles of the \ion{Ca}{2} H\&K lines from the
high-resolution spectra.  Zero velocity is defined relative to the
center of the narrow absorption component, and corresponds to a
redshift of 170 km s$^{-1}$.  The absorption features are broad and
asymmetric, extending from about $-400$ to $+1100$ km s$^{-1}$ on June
6 UT.  The width of the red wing decreases with time, and the blue
wing turns to emission.  The dashed red and blue lines are,
respectively, Lorentzian and Gaussian fits to the red wing of the
lines.
\label{fig:cahk}}
\end{figure}

\begin{figure}
\epsscale{1}
\plotone{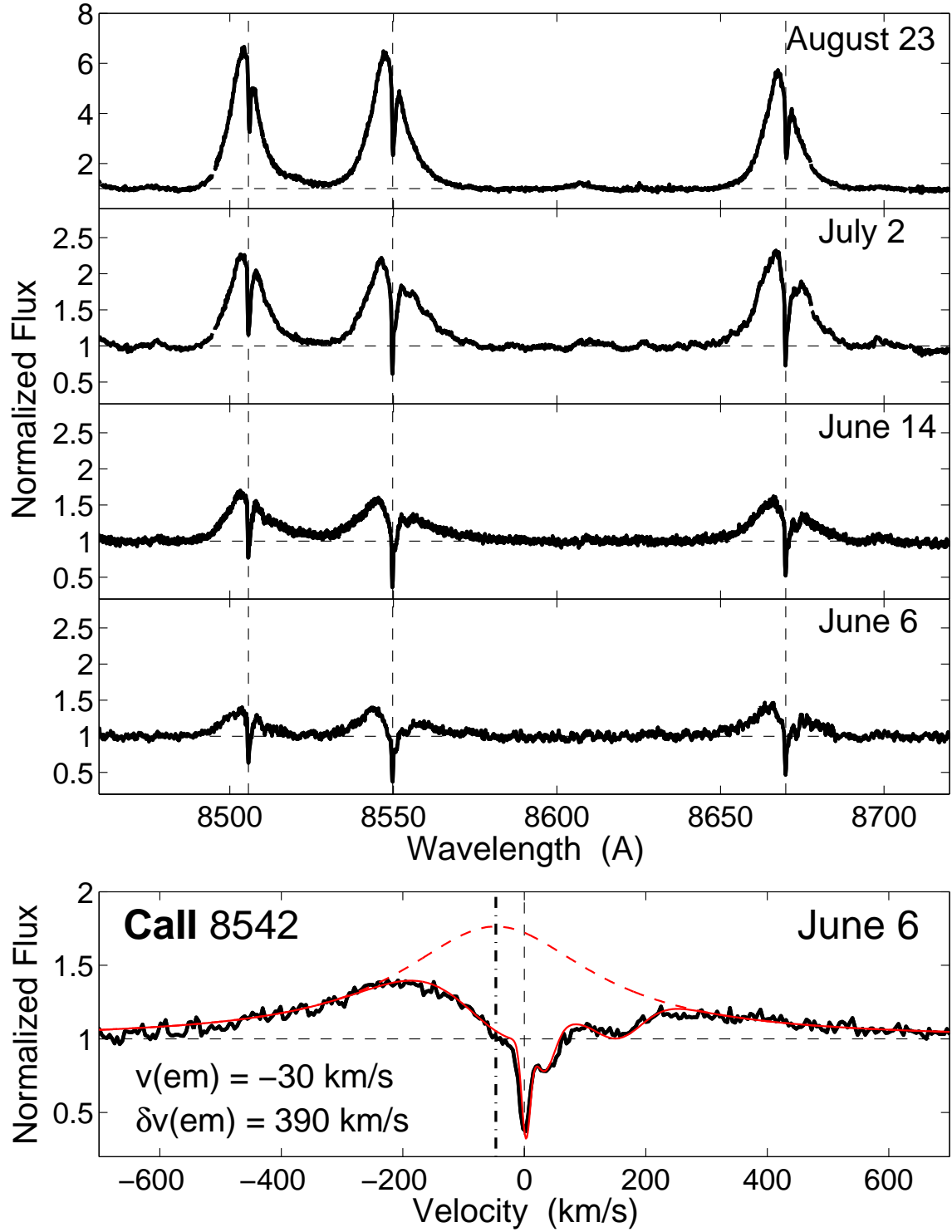}
\caption{{\it Top:} Profiles of the \ion{Ca}{2} IR triplet lines from
the high-resolution spectra.  {\it Bottom:} Zoom-in on the
\ion{Ca}{2}$\,8542$ line from June 6.  The solid red line is a
combined model that includes a Lorentzian emission line (dashed red
line; peak marked by dot-dashed blue line), absorbed by 4 distinct
components: an interstellar unresolved component defining the zero
velocity, and three broader absorption components that are all
centered at velocities redshifted relative to the emission line
center.  The emission line center is blueshifted by about 50 km
s$^{-1}$ relative to the interstellar absorption.
\label{fig:cair}}
\end{figure}

\begin{figure}
\epsscale{1}
\plotone{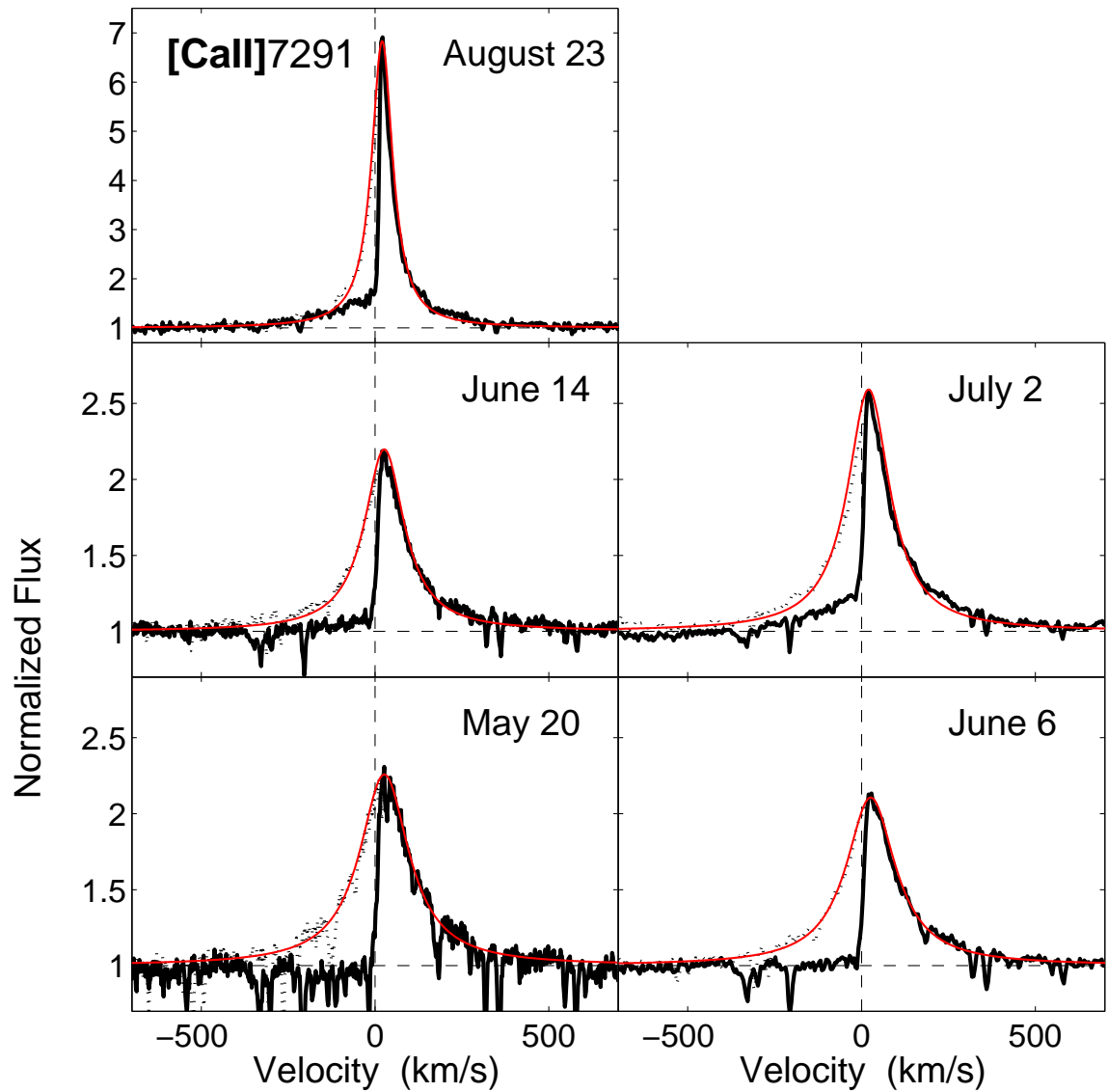}
\caption{Emission lines of [\ion{Ca}{2}]$\,7291$ from the
high-resolution spectra.  The lines are strongly asymmetric, with the
blue wing of the line completely missing.  A low level of blue-wing
emission emerges on July 2, but it does not appear to brighten to the
level of the red wing by August 23.  The line width clearly decreases
with time.  The dotted black lines are a reflection of the red wing.
The red lines are Lorentzian profile fits.  The narrow absorption
features in the various spectra are due to telluric lines.
\label{fig:caii_1}}
\end{figure}

\begin{figure}
\epsscale{1}
\plotone{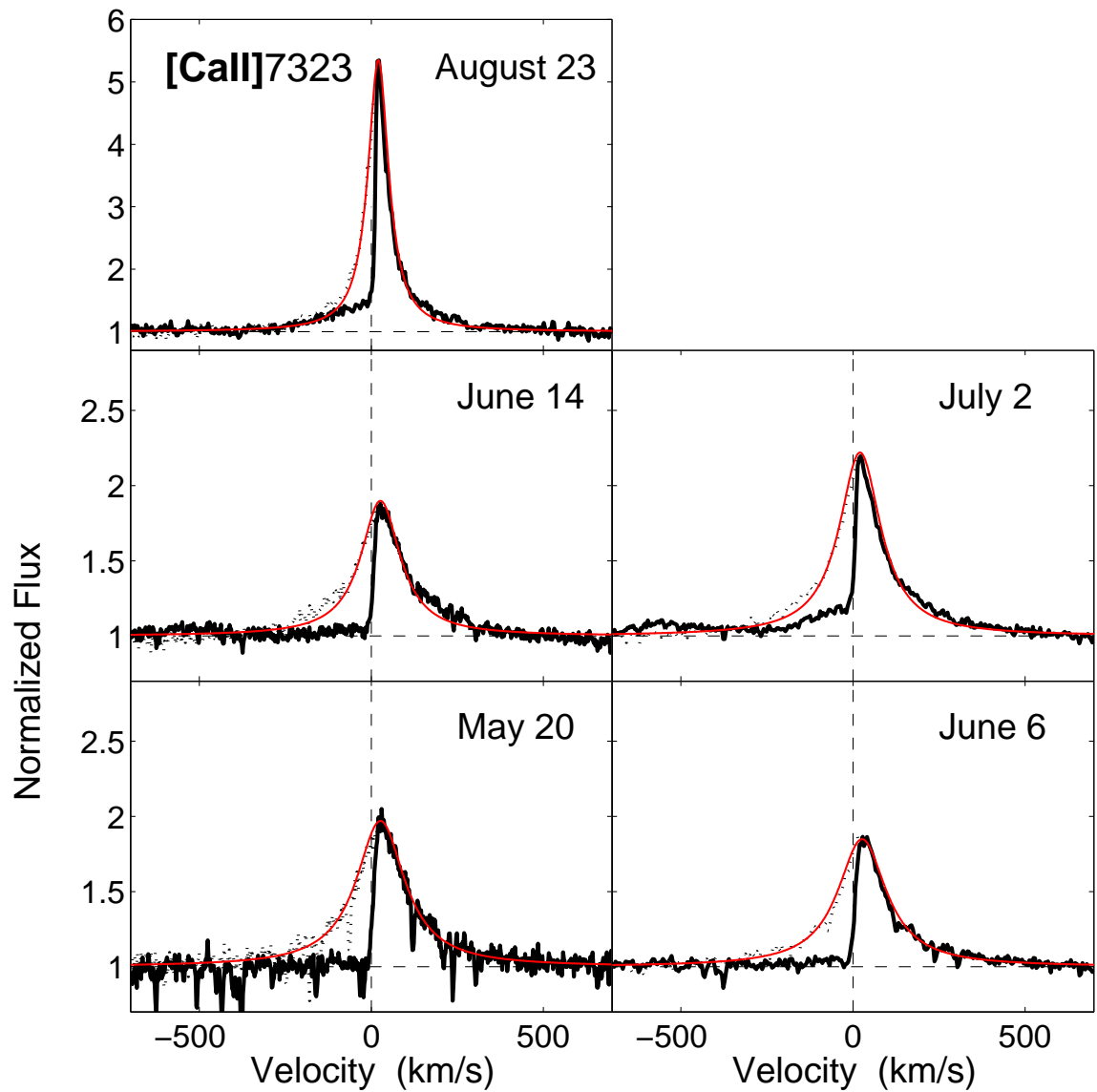}
\caption{Same as Figure~\ref{fig:caii_1} but for the weaker 
[\ion{Ca}{2}]$\,7323$ line.
\label{fig:caii_2}}
\end{figure}

\begin{figure}
\epsscale{1}
\plotone{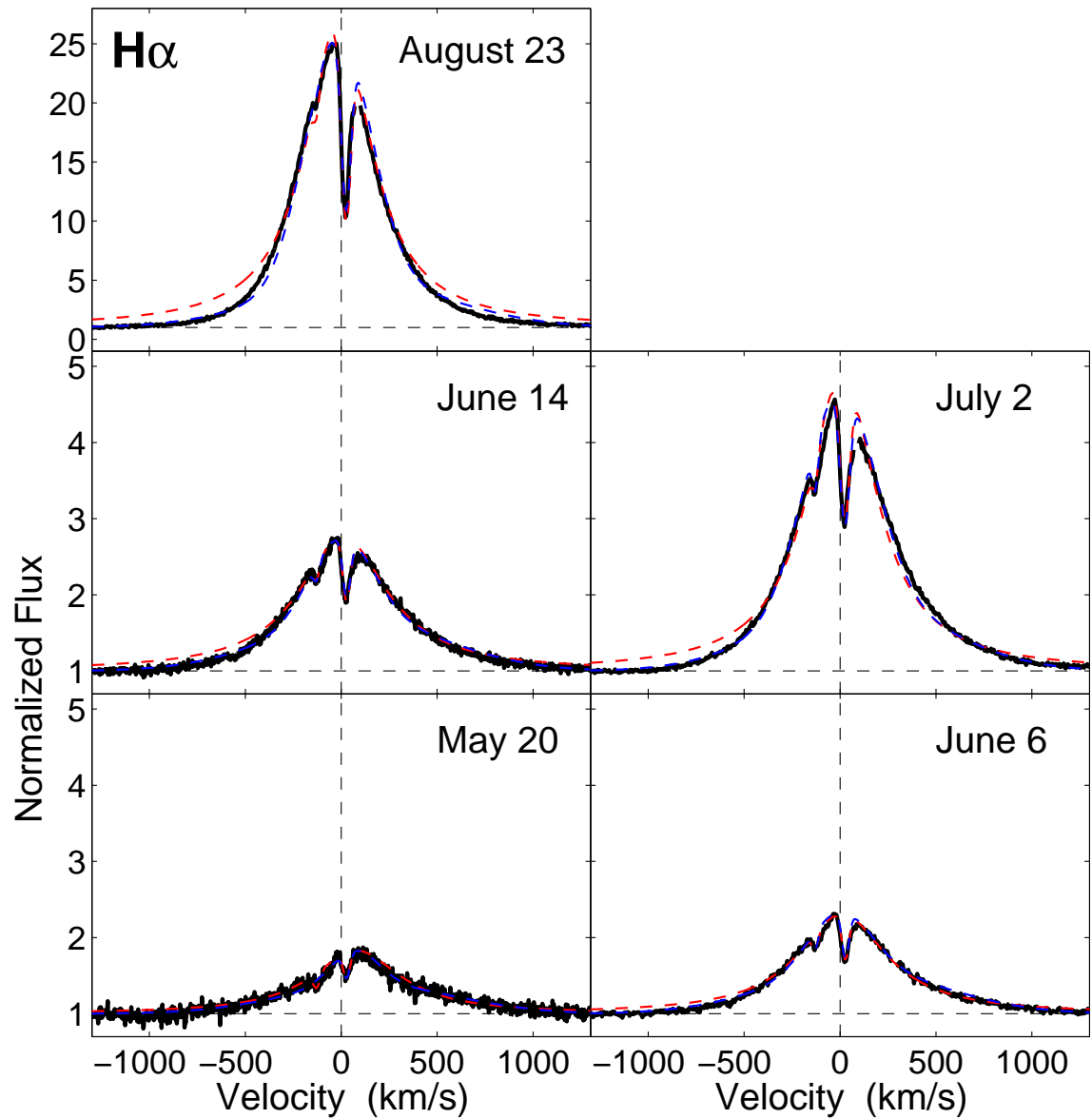}
\caption{H$\alpha$ emission lines from the high-resolution spectra.
The lines are overall symmetric but are marked by a deep narrow
absorption feature redward of the line center (a second absorption
feature is possibly detected blueward of the line center).  The dashed
red lines are Lorentzian model fits, while the dashed blue lines are
models with offset narrow and wide Gaussian components.
\label{fig:ha}}
\end{figure}

\begin{figure}
\epsscale{1}
\plotone{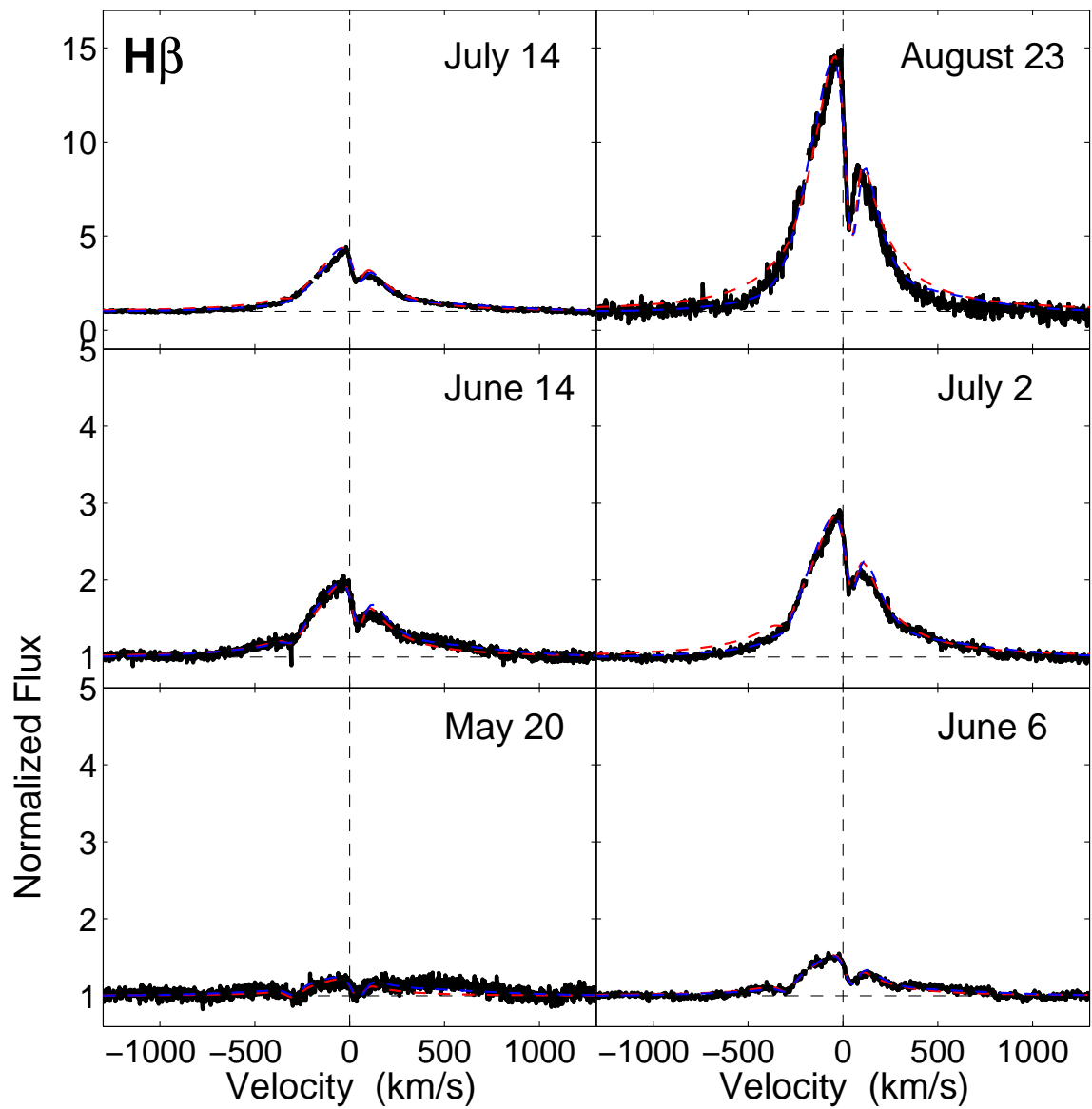}
\caption{Same as Figure~\ref{fig:ha} but for the H$\beta$ emission 
line. 
\label{fig:hb}}
\end{figure}

\begin{figure}
\epsscale{1}
\plotone{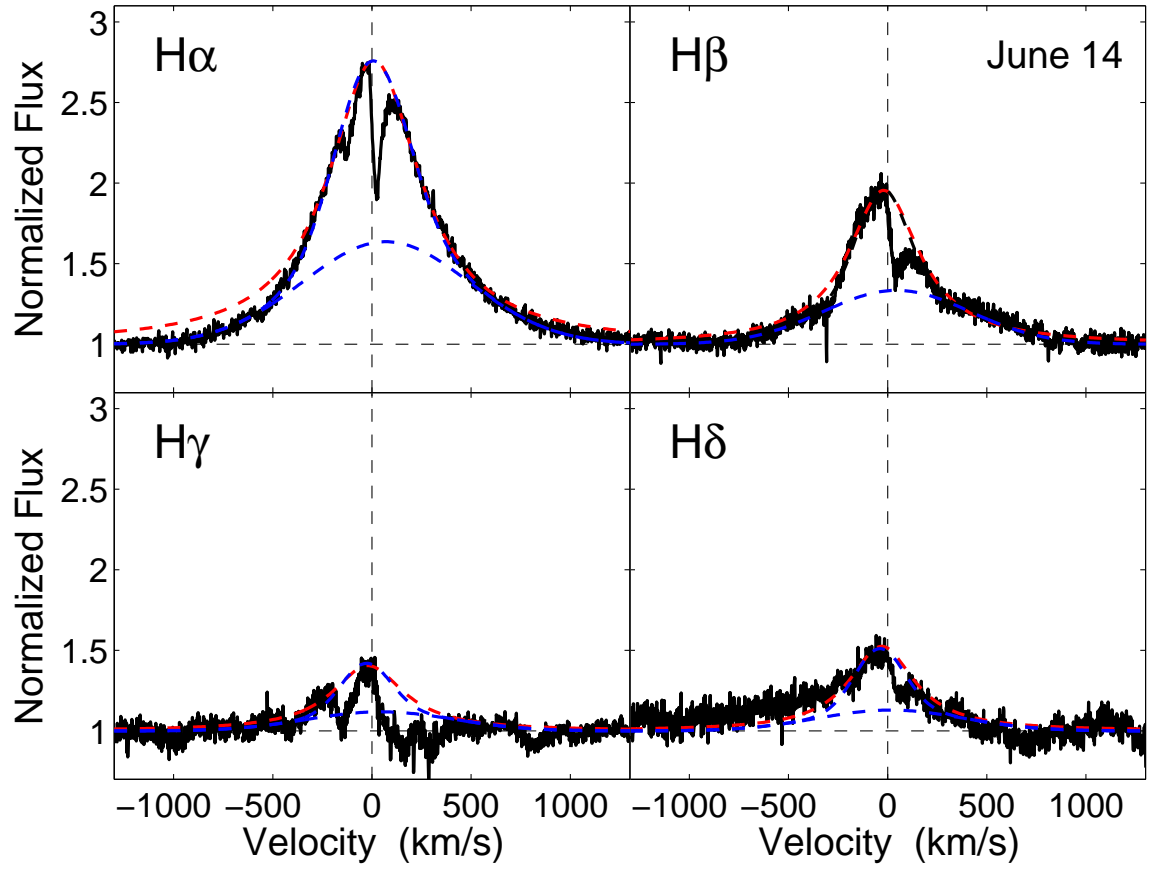}
\caption{Hydrogen Balmer lines from the high resolution spectrum on
June 14.  The red and blue dashed lines are the same as in
Figure~\ref{fig:ha}, with the broad Gaussian component shown
separately.  We do not include the absorption features in these fits.
\label{fig:balmer3}}
\end{figure}

\begin{figure}
\epsscale{1}
\plotone{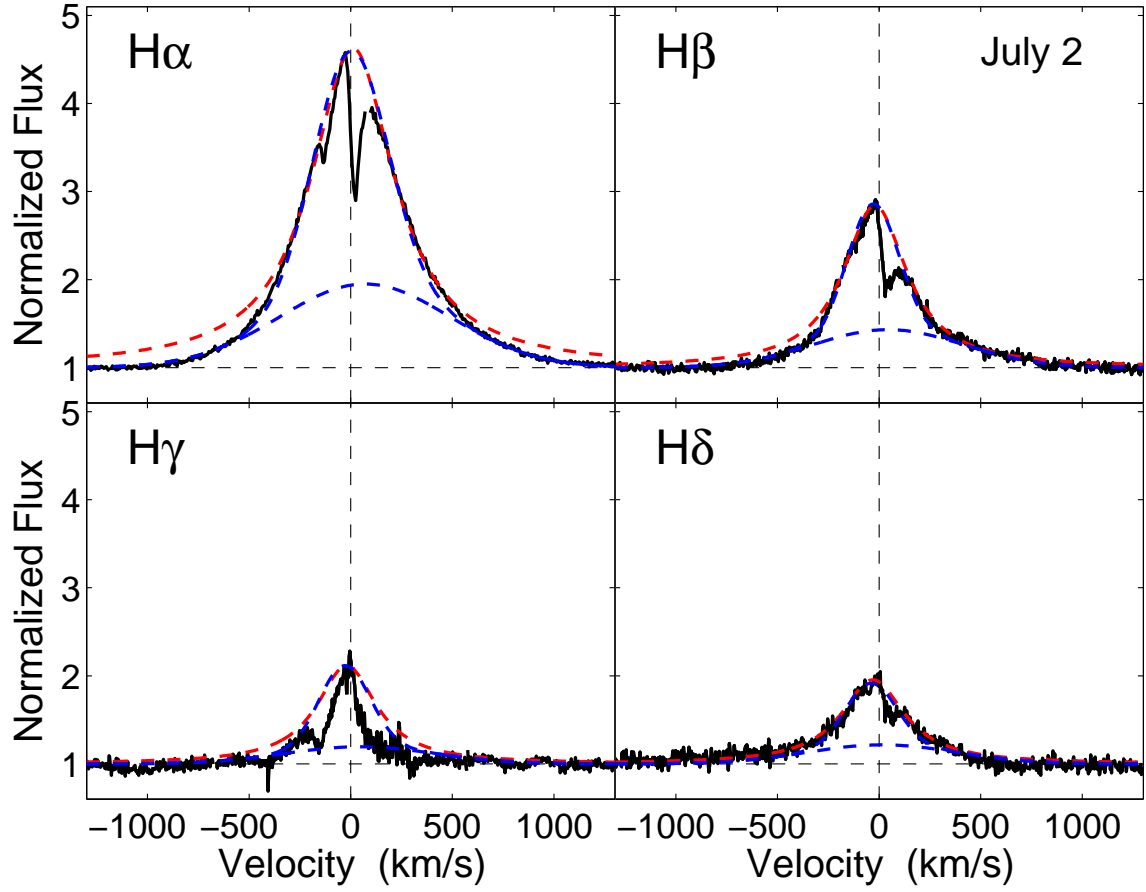}
\caption{Same as Figure~\ref{fig:balmer3} but for the spectrum from
July 2.
\label{fig:balmer4}}
\end{figure}

\begin{figure}
\epsscale{1}
\plotone{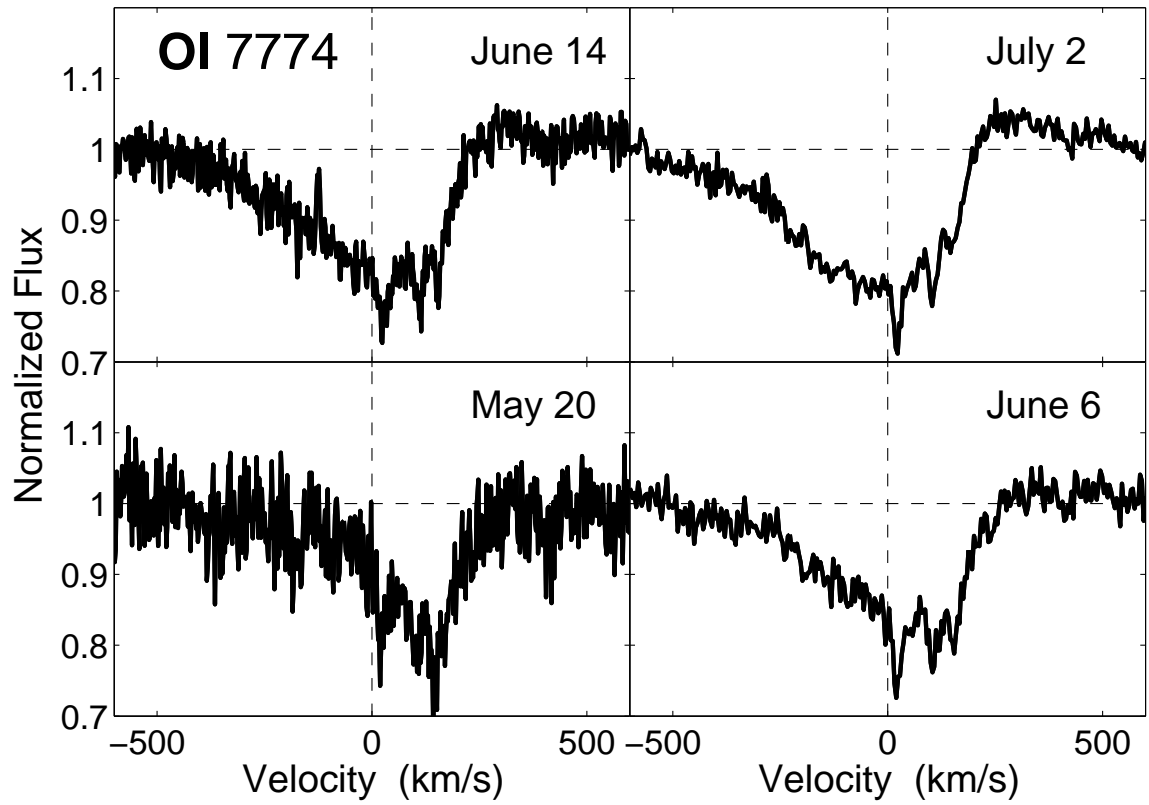}
\caption{\ion{O}{1}$\,\lambda 7774$ triplet absorption profiles from
the high-resolution spectra.  The profiles are a combination of narrow
interstellar components (clearly seen in the June 6, June 14, and July
2 spectra), and a broad component that is likely circumstellar in
origin.
\label{fig:oi}}
\end{figure}

\begin{figure}
\epsscale{1}
\plotone{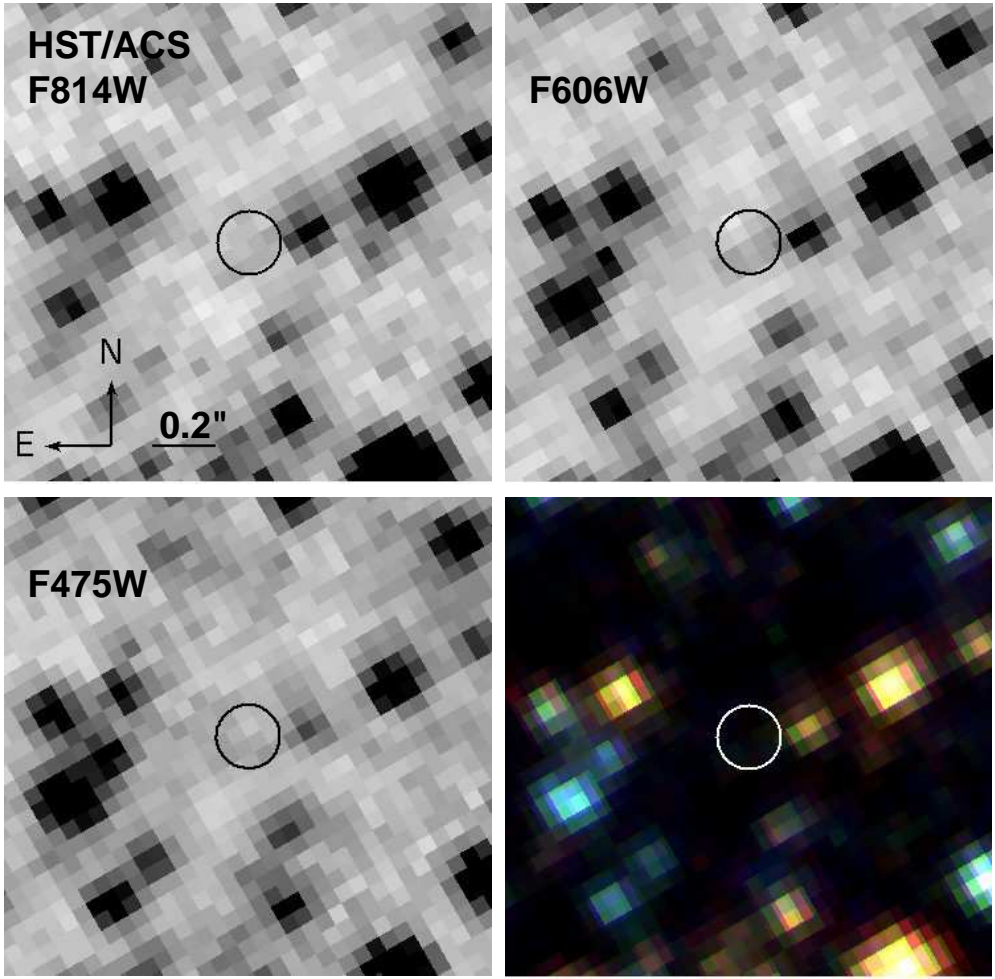}
\caption{HST/ACS images at the location of \trans\ revealing no
coincident source.
\label{fig:hst}} 
\end{figure}

\begin{figure}
\epsscale{1}
\plotone{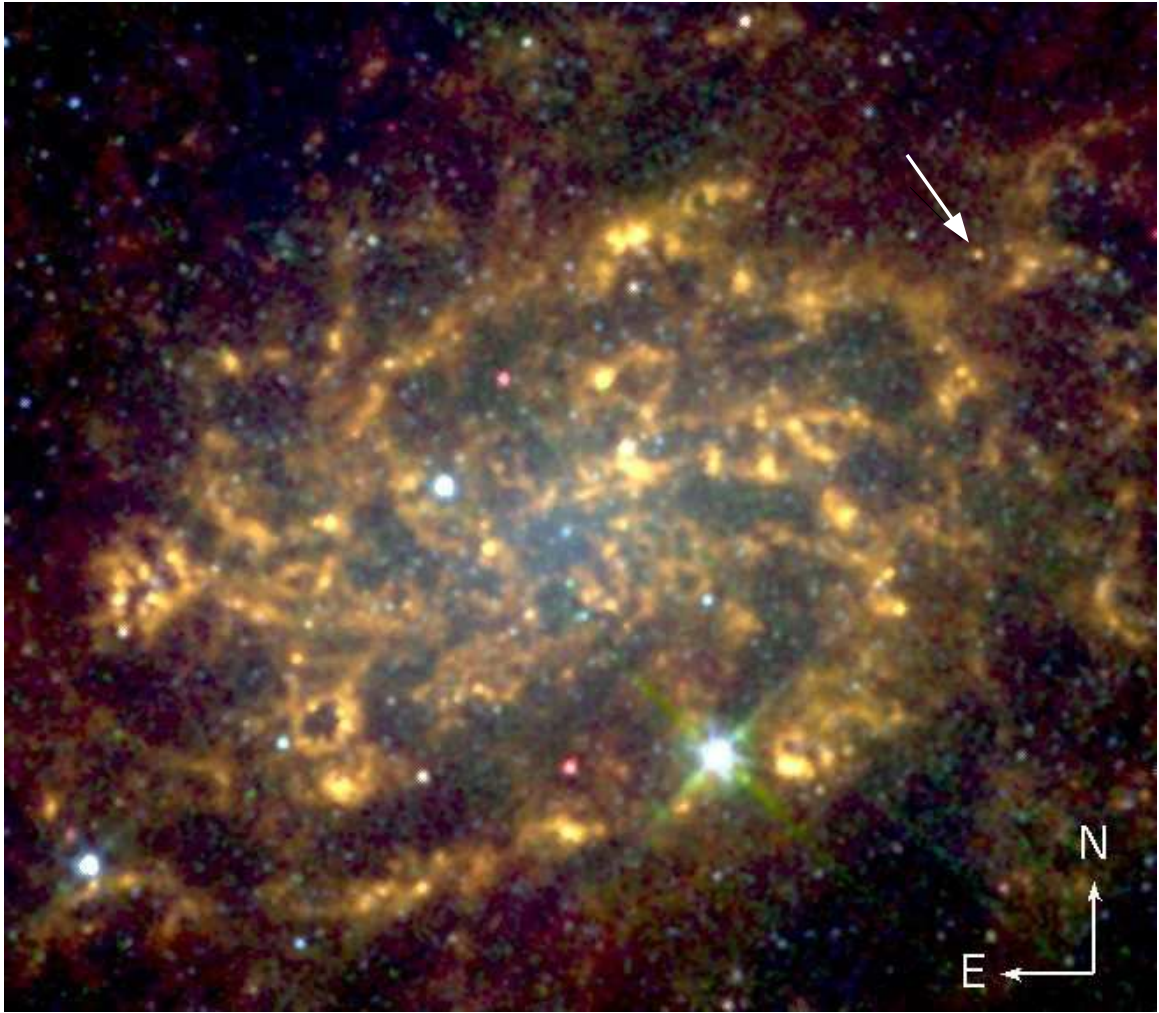}
\caption{{\it Spitzer} false-color image of NGC\,300 with the location
of \trans\ marked by an arrow.  The red, green, and blue channels
correspond to the 4.5, 5.8 and 8.0 $\mu$m IRAC images, respectively.
\label{fig:spitzer1}} 
\end{figure}

\begin{figure}
\epsscale{1}
\plotone{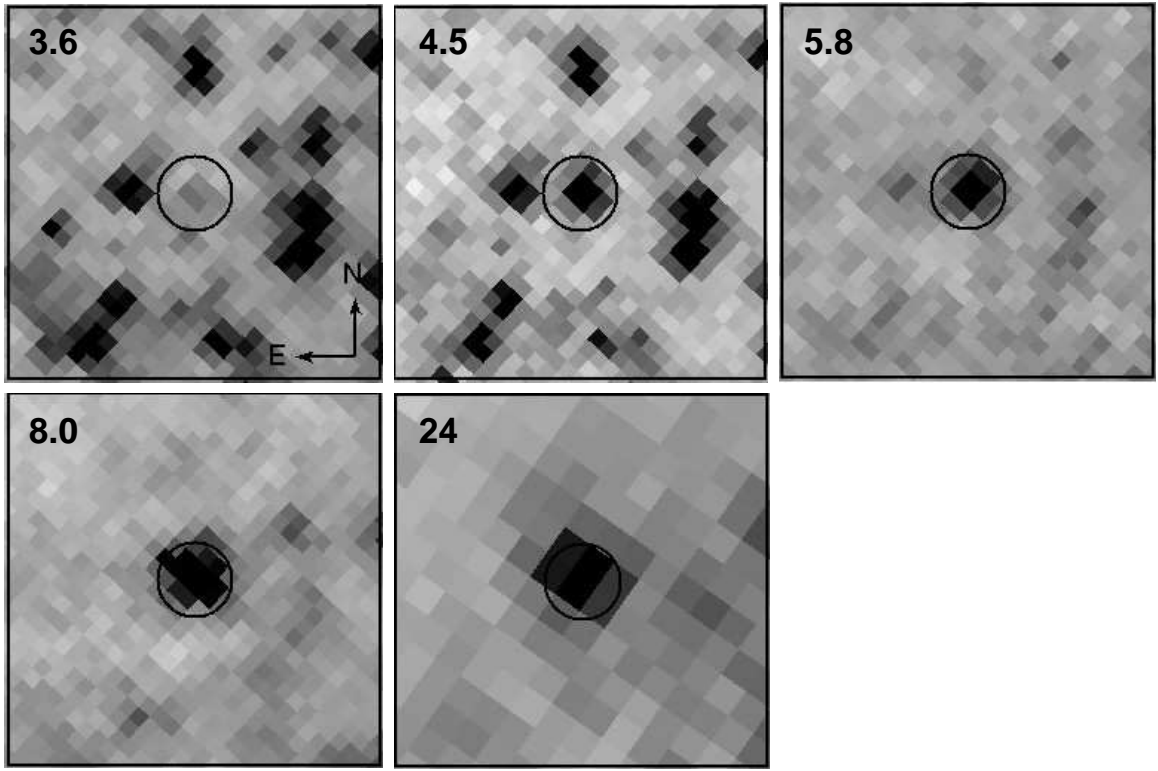}
\caption{{\it Spitzer} IRAC and MIPS images of the progenitor of
\trans\ from 3.6 to 24 $\mu$m.
\label{fig:spitzer2}} 
\end{figure}

We finally turn to a comparison of \trans\ with two recent events that
exhibit several similar properties: \m85\ and SN\,2008S.  SN\,2008S
was discovered on February 1.8 UT in NGC\,6946 ($d\approx 5.6$ Mpc)
and had an absolute magnitude of $M_V\approx -14$ mag (taking into
account significant Galactic and host obscuration of $A_V\approx 2.5$
mag; \citealt{pkt+08,sgl+08}).  The $R$-band light curve of SN\,2008S
from \citet{sgl+08} is shown in Figure~\ref{fig:lcs}.  The temporal
evolution is remarkably similar to that of \trans, namely a relatively
stable decay rate of about 0.035 mag d$^{-1}$ from about 30 to 100
days.  However, unlike \trans, the optical emission from SN\,2008S
does not exhibit a clear steepening at later times, and it may indeed
flatten to a decay rate of about 0.01 mag d$^{-1}$ from 130 to 270
days.

\begin{figure}
\epsscale{1}
\plotone{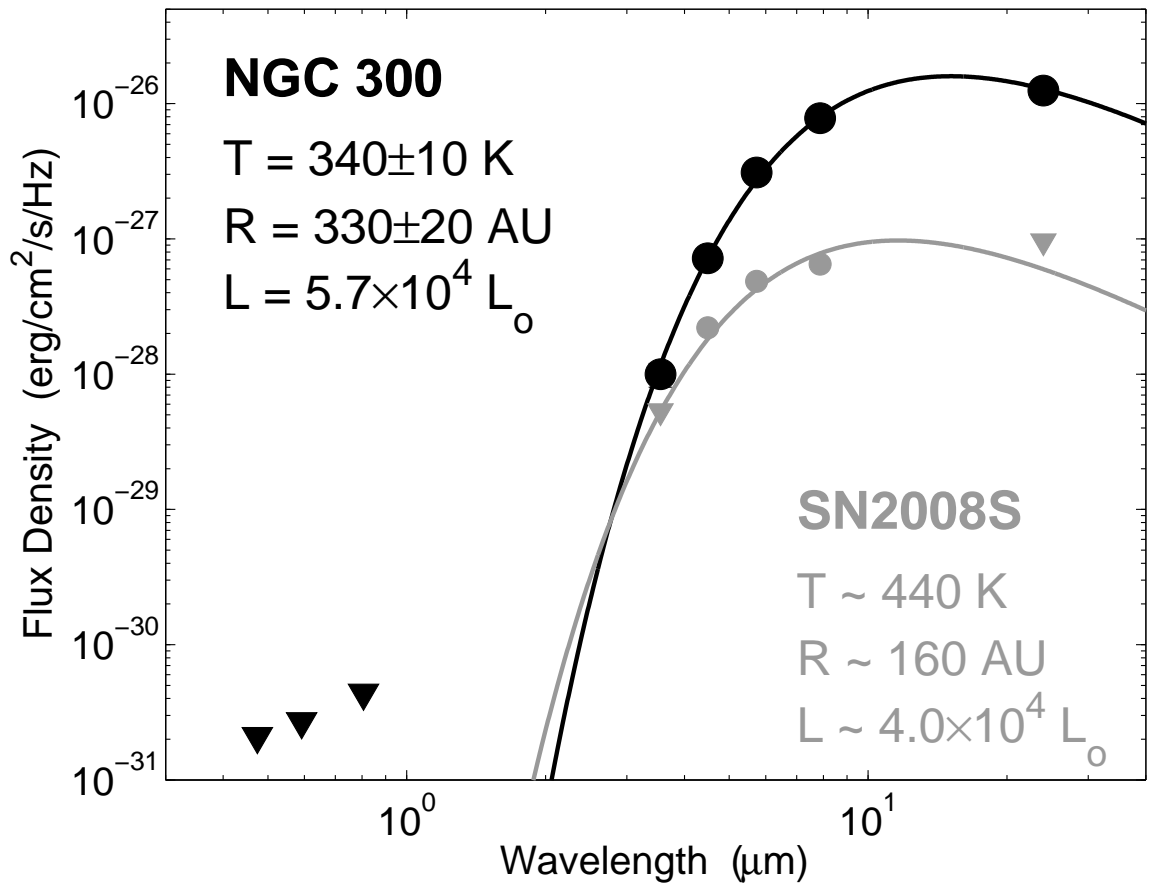}
\caption{Spectral energy distribution of the archival source
coincident with \trans\ from HST (upper limits) and {\it Spitzer}
(black circles) observations.  The black line is the best-fit
blackbody model ($\chi^2_r=1.5$ for 3 degrees of freedom).  Also shown
is the SED of the {\it Spitzer} source coincident with SN\,2008S
\citep{pkt+08,tps+08} and its best-fit blackbody model.  The two
progenitors share similar properties.
\label{fig:spitzer_hst}} 
\end{figure}

\begin{figure}
\epsscale{1}
\plotone{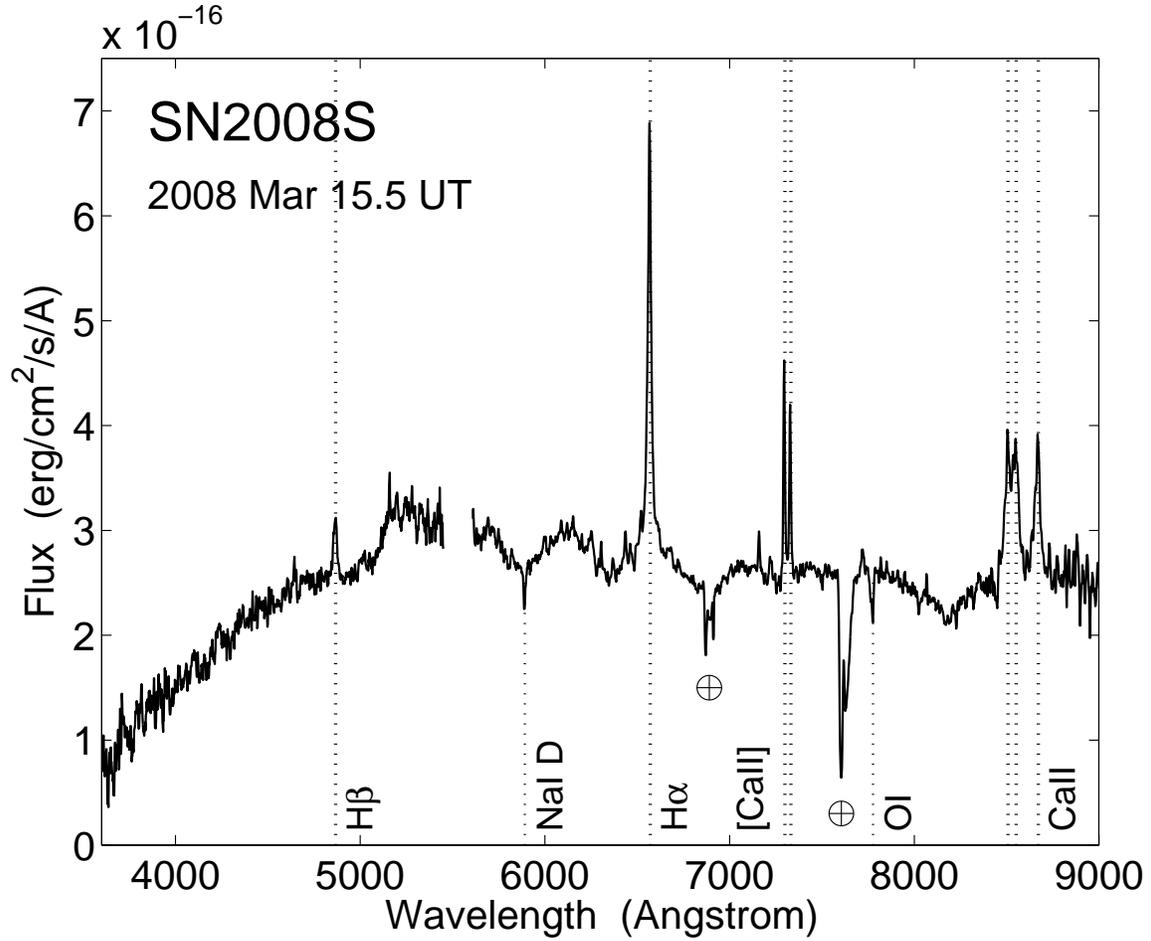}
\caption{Low-resolution spectrum of SN\,2008S obtained with the ARC
3.5-m telescope at Apache Point Observatory about 42 d after
discovery.  The continuum shape and brightness, as well as the
observed spectral features closely resemble those of \trans,
indicating that the two events likely share a common origin.  We note,
however, the lack of \ion{Ca}{2} H\&K absorption and \ion{He}{1}
emission, which are prominent in the spectrum of \trans.
\label{fig:2008s}} 
\end{figure}

In Figure~\ref{fig:2008s} we present a low-resolution spectrum of
SN\,2008S that we obtained on 2008 Mar 15.5 UT ($\delta t\approx 42$
d) using the ARC 3.5-m telescope at Apache Point Observatory.  The
resolution of the spectrum is 6.5 \AA, or about 300 km s$^{-1}$.
Neglecting extinction, we find that the overall continuum shape and
brightness are characterized by a blackbody profile with $T\approx
4800$ K and $R\approx 1.6\times 10^{14}$ cm, similar to the properties
inferred for \trans\ at a similar epoch of about 44 d
(Figure~\ref{fig:spitzer_hst}).  The inferred bolometric absolute
magnitude is $\approx -11.3$ mag.  However, if we include the
estimated extinction of $A_V\approx 2.5$ mag, we find that the
spectrum requires a higher temperature, $T\approx 10^4$ K, and a
smaller radius, $R\approx 9\times 10^{13}$ cm, leading to an overall
absolute magnitude of $M_{\rm bol}\approx -13.9$ mag, similar to the
values inferred by \citet{sgl+08} from similar spectroscopic
observations.  Thus, SN\,2008S has a bolometric luminosity that is
nearly an order of magnitude larger than \trans.  Combined with the
similar temporal evolution, we conclude that SN\,2008S released about
an order of magnitude more energy than \trans, $E\sim 10^{48}$ erg
(see also \citealt{sgl+08}).

The overall similarity in also evident in the dominant spectral
features of both events.  SN\,2008S exhibits strong and narrow
hydrogen Balmer lines, [\ion{Ca}{2}] lines, and \ion{Ca}{2} IR lines.
The H$\alpha$ line is resolved with a width of $\delta v_{\rm
fwhm}\approx 1200$ km s$^{-1}$, about a factor of 2 times broader than
for \trans.  As in the case of \trans, the \ion{Ca}{2} IR triplet
lines exhibit a similar width to that of H$\alpha$, $\delta v_{\rm
fwhm}\approx 1300$ km s$^{-1}$.  The [\ion{Ca}{2}] lines are only
marginally resolved, with $\delta v_{\rm fwhm}\approx 350$ km
s$^{-1}$.  This is again a factor of 2 times broader than in \trans,
and comparably narrower than the H$\alpha$ line.  Due to the low
resolution of the spectrum it is unclear whether the [\ion{Ca}{2}]
lines are asymmetric.  Unlike in the case of \trans, we do not detect
\ion{Ca}{2} H\&K absorption or \ion{He}{1} emission.

Finally, perhaps the most important similarity between SN\,2008S and
\trans\ is the detection of a coincident {\it Spitzer} source in both
events \citep{pkt+08,tps+08}.  As shown in
Figure~\ref{fig:spitzer_hst}, the SEDs of both coincident sources are
similar, with $T\approx 440$ K, $R\approx 160$ AU, and $L\approx
4\times 10^{4}$ L$_\odot$ for SN\,2008S.  Thus, both events appear to
arise from objects of a similar luminosity and in a similar
evolutionary stage.  \citet{tps+08} recently argued that the
progenitors of both events are extreme AGB stars with masses of $\sim
10$ M$_\odot$, and that the events themselves are likely
electron-capture SNe.  \citet{sgl+08}, on the other hand, advocated a
scenario for SN\,2008S of a super-Eddington eruption that results in
significant mass loss.  They further note an overall similarity to
IRC+10420, and speculate that SN\,2008S resembles an eruption of a
$\sim 20$ M$_\odot$ blue supergiant, rather than a lower mass red
supergiant as advocated by \citet{tps+08}.  Our high resolution
spectroscopic observations of \trans, and their close resemblance to
the spectrum of IRC+10420, support the interpretation of
\citet{sgl+08}.  Still, while both events appear to arise from similar
objects, these progenitors are clearly capable of producing eruptions
with an energy release that spans at least an order of magnitude.

We next turn to a comparison with \m85.  As shown in
Figure~\ref{fig:optsed}, the spectral energy distribution of this
event closely resembles \trans, with a nearly identical bolometric
luminosity of about $-12$ mag \citep{kor+07}.  The light curve
evolution is also similar (Figure~\ref{fig:lcs}), with a comparable
decay rate during the first $\sim 70$ days, followed by a significant
steepening at later time.  Thus, \m85\ appears to be a clear analog of
both \trans\ and SN\,2008S.

Unfortunately, the low-resolution spectra of \m85, obtained about 1
and 48 d after discovery \citep{kor+07}, had a low signal-to-noise
ratio that precludes a detailed analysis.  Still, the second spectrum
does exhibit a narrow H$\alpha$ emission line with $\delta v_{\rm
fwhm}\approx 350$ km s$^{-1}$, about two times narrower than in
\trans.  Additional weak features possibly corresponding to
\ion{Ca}{2} H\&K and \ion{Fe}{2}$\lambda 6544$ may also be discerned
in the spectrum (see Figure 3 of \citealt{kor+07}).  These possible
features provide additional support for a common origin with \trans.

On the other hand, the environments of \trans\ and \m85\ differ
significantly.  In the case of \m85, the host is an early type galaxy
(with a mean age of $\sim 1.6$ Gyr), and the environment of the
transient itself contains no stars brighter than $M_z\approx -6.2$
mag, corresponding to a mass of $\lesssim 7$ M$_\odot$ \citep{okr+08}.
The local environment of \trans, on the other hand, contains stars
with inferred masses of $\sim 14$ M$_\odot$ \citep{bbh+09}, in line
with the inferred mass of the progenitor system.  Thus, it appears
that different channels/progenitors may lead to the formation of
similar transient outbursts.

Given the overall similarity between the three events it is reasonable
to conclude that they share a common origin.  Thanks to our detailed
high-resolution spectroscopic observations of \trans, an intimate
connection with objects like IRC+10420 appears likely (in agreement
with previous speculation \citealt{sgl+08}) but may not be required
(e.g., \m85).  From the overall similarity we further conclude that
none of the three transients are likely to have resulted from the
explosion of the progenitor star.  Instead, the eruptions may mark a
shift in the evolutionary stage from complete obscuration to a yellow
supergiant phase, and eventually a Wolf-Rayet star.  Clearly, such a
transition apparently leads to overall common features, but in detail
the energy and velocity of the outflow appear to vary by at least an
order of magnitude.

\section{Summary and Conclusions} 
\label{sec:conc}

We presented the first high-resolution spectroscopic observations for
the intermediate luminosity optical transient \trans, as well as
multi-wavelength observations (UV/optical/radio/X-ray) and an
independent analysis of archival {\it Spitzer} and HST observations of
its progenitor (see also \citealt{tps+08,bbh+09}).  The preponderance
of photometric and spectroscopic evidence indicate that this event is
not a supernova, but is instead the result of an eruption on a $\sim
10-20$ M$_\odot$ blue supergiant or pre-Wolf-Rayet star, possibly in a
binary system.  The high-resolution spectra reveal the presence of a
complex circumstellar environment, marked by outflows and possible
inflow from previous ejections or from a companion wind.  The observed
inflow velocities are indicative of a massive and relatively compact
progenitor, not a highly extended red supergiant, in contrast to the
interpretation of \citet{tps+08} and \citet{bbh+09}.  Alternatively,
the progenitor may be part of a binary system in which the companion
wind (with a velocity of $\sim 10^3$ km s$^{-1}$) is observed in
absorption along the line of sight.  In this scenario, however, it may
be difficult to accommodate the relatively low total system mass of
$\lesssim 20$ M$_\odot$.

The outburst itself gave rise to an outflow with a velocity of $\sim
600$ km s$^{-1}$ and a kinetic energy of at least $10^{47}$ erg.  No
radio or X-ray emission are detected to limits that are significantly
dimmer than the faintest SNe detected to date, indicating a lack of
fast ejecta (with $v\gtrsim 0.1c$).

\trans\ closely resembles two previous events, \m85\ and SN\,2008S, in
terms of its intermediate luminosity and spectral properties.  It also
shares with SN\,2008S a dust-obscured progenitor with a luminosity of
a $\sim 10^5$ L$_\odot$.  Our detailed comparison reveals that the
outburst and progenitor properties point to a common origin, but with
a range of outburst energy and velocity of nearly an order of
magnitude.  We also note a resemblance between the spectroscopic
properties of \trans\ and those of the luminous yellow supergiant
IRC+10420.  However, the latter is about an order of magnitude more
luminous than the progenitor of \trans, and its observed ``quiescent''
outflow has a velocity of only $\sim 50$ km s$^{-1}$.  These
differences raise the possibility that \trans\ marks the transition to
an IRC+10420-like state, or that its progenitor is simply a lower mass
analogue.

While our optical/UV photometry and low-resolution spectroscopy are of
similar quality to those of SN\,2008S and \m85, our extensive echelle
spectroscopy provides a much deeper view of the circumstellar
environment than previously available for this class of events.
Indeed, we stress that the overall photometric behavior and general
spectroscopic features observed in \trans\ may be generic to a broad
class of stellar eruptions with an intermediate velocity in the
presence of a dense circumstellar environment.  For example, the 2002
eruption of V838 Mon was similar in broad terms to that of \trans,
\m85, and SN\,2008S, based on a comparable light curve behavior and
spectroscopic evidence for narrow H$\alpha$ and \ion{Ca}{2} lines
(e.g., \citealt{rgf+05}).  However, the eruption was about an order of
magnitude less luminous than \trans\ and \m85, and the progenitor
appears to have been a lower mass star.  We conclude that a clear
understanding of the broad range of ILOTs requires high-resolution
spectroscopy to assess the detailed properties of the circumstellar
environment, as well as the identification of progenitors; optical/IR
photometry and low-resolution spectroscopy alone are not likely to be
sufficient.

Finally, the main prediction of our model is that the progenitor of
\trans\ should be visible following the decay of the event.  If the
current eruption is similar to previous events that produced the
now-destroyed dust shell, than the progenitor would again be visible
as an obscured IR source.  However, if only a minor amount of dust
reforms, we predict that the spectrum will more closely resemble that
of IRC+10420.  In either case it appears that events like \trans\
represent an important stage in the evolution of massive stars toward
their eventual demise in SN explosions.  The advent of facilities such
as Pan-STARRS and LSST will lead to an increased rate of detections of
events similar to \trans.  From the observations presented here it is
clear that multi-wavelength follow-up, in particular high-resolution
spectroscopy, which is not typically attempted for SNe
(cf.~SN\,2008S), is critical for a detailed understanding of the
eruption process and the properties of the progenitor(s).

\acknowledgements This paper includes data gathered with the 6.5 meter Magellan Telescopes
located at Las Campanas Observatory, Chile.  We thank David Sand for obtaining the LDSS-3
spectrum on 2008 June 5, Bruce Draine for insightful discussions about the spectral properties
of \trans, and Ranga Chary for assistance with the {\it Spitzer} data.  AMS acknowledges support
by NASA through a Hubble Fellowship grant.  RAC acknowledges support from an NSF grant
AST-0807727.


\clearpage
\begin{deluxetable}{llccccc}
\tabletypesize{\footnotesize}
\tablecolumns{7}
\tabcolsep0in\footnotesize
\tablewidth{0pc}
\tablecaption{Spectroscopic Observations \label{tab:spec}}
\tablehead {
\colhead {UT Date}          &
\colhead {Instrument}       &
\colhead {Exposure}         &
\colhead {Grism}            &
\colhead {Slit}             &
\colhead {Wavelength}       &
\colhead {Resolution}       \\
\colhead {}                 &
\colhead {}                 &
\colhead {(s)}              &
\colhead {}                 &
\colhead {}                 &
\colhead {(\AA)}            &    
\colhead {(\AA)}                
}
\startdata
2008 May 20.42    & MIKE  & $2\times 600$  & Blue     & $0.7''$ & $3400-5100$ & 0.10  \\
                  & MIKE  & $2\times 600$  & Red      & $0.7''$ & $4900-9400$ & 0.21  \\
2008 May 23.43    & LDSS-3 & $1\times 900$  & VPH-red  & $1''$   & $6000-9700$ & 6.2  \\
2008 May 31.40    & LDSS-3 & $2\times 600$  & VPH-blue & $1''$   & $3800-6500$ & 3.4  \\
                  & LDSS-3 & $2\times 720$  & VPH-red  & $1''$   & $6000-9700$ & 6.2  \\
2008 June 5.41    & LDSS-3 & $1\times 900$  & VPH-red  & $1''$   & $6000-9700$ & 6.2  \\
2008 June 6.40    & MIKE  & $3\times 600$  & Blue     & $1''$   & $3400-5100$ & 0.13 \\  
                  & MIKE  & $3\times 600$  & Red      & $1''$   & $4900-9400$ & 0.28 \\ 
2008 June 14.40   & MIKE  & $2\times 1200$ & Blue     & $0.7''$   & $3400-5100$ & 0.10 \\  
                  & MIKE  & $2\times 1200$ & Red      & $0.7''$   & $4900-9400$ & 0.21 \\ 
2008 June 22.42   & LDSS-3 & $1\times 900$  & VPH-blue & $1''$   & $3850-6600$ & 3.4  \\
                  & LDSS-3 & $1\times 900$  & VPH-red  & $1''$   & $6050-9750$ & 6.2  \\
2008 July 2.35    & MIKE  & $2\times 900$  & Blue     & $1''$   & $3400-5100$ & 0.13 \\  
                  & MIKE  & $2\times 900$  & Red      & $1''$   & $4900-9400$ & 0.28 \\ 
2008 July 14.36   & MIKE  & $2\times 900$  & Blue     & $1''$   & $3400-5100$ & 0.13 \\  
                  & MIKE  & $2\times 900$  & Red      & $1''$   & $4900-9400$ & 0.28 \\ 
2008 August 23.32 & MIKE  & $2\times 900$  & Blue     & $0.7''$ & $3400-5100$ & 0.10 \\ 
                  & MIKE  & $2\times 900$  & Red      & $0.7''$ & $4900-9400$ & 0.21 \\ 
\enddata
\tablecomments{Journal of spectroscopic observations of the transient
in NGC\,300.}
\end{deluxetable}

\clearpage
\begin{deluxetable}{lcccc}
\tabletypesize{\footnotesize}
\tablecolumns{5}
\tabcolsep0in\footnotesize
\tablewidth{0pc}
\tablecaption{{\it Swift} UV/Optical Telescope Observations \label{tab:uvot}}
\tablehead {
\colhead {UT Date}          &
\colhead {Filter}           &
\colhead {Exposure}          &
\colhead {Magnitude}        &
\colhead {Flux Density}     \\
\colhead {}                 &
\colhead {}                 &
\colhead {(s)}              &
\colhead {}                 &
\colhead {($10^{-15}$ erg s$^{-1}$ cm$^{-2}$ \AA$^{-1}$)}
}
\startdata
2008 May 20.01     & UVW2 & 5723 & $18.24\pm 0.03$ & $0.27\pm 0.01$ \\
2008 May 25.03     & UVW1 & 3877 & $17.67\pm 0.03$ & $0.34\pm 0.01$ \\
2008 May 25.04     & U    & 1551 & $16.13\pm 0.01$ & $1.25\pm 0.01$ \\
2008 May 25.04     & B    & 773  & $15.68\pm 0.01$ & $3.46\pm 0.03$ \\
2008 May 25.04     & V    & 691  & $14.88\pm 0.02$ & $4.18\pm 0.06$ \\
2008 June 4.68     & UVW1 & 1295 & $17.90\pm 0.05$ & $0.27\pm 0.01$ \\
2008 June 4.67     & U    & 1599 & $16.43\pm 0.01$ & $0.94\pm 0.01$ \\
2008 June 4.68     & B    & 1856 & $15.93\pm 0.01$ & $2.75\pm 0.02$ \\
2008 June 4.67     & V    & 2006 & $15.02\pm 0.02$ & $3.69\pm 0.05$ \\
2008 June 19.39    & UVW1 & 1765 & $17.96\pm 0.05$ & $0.26\pm 0.01$ \\
2008 June 19.39    & U    & 879  & $16.70\pm 0.02$ & $0.74\pm 0.01$ \\
2008 June 19.39    & B    & 879  & $16.22\pm 0.02$ & $2.11\pm 0.03$ \\
2008 June 19.39    & V    & 879  & $15.21\pm 0.02$ & $3.08\pm 0.05$ \\
2008 August 14.17  & UVW1 & 746  & $>20.6$         & $<0.02$        \\
2008 August 14.17  & U    & 428  & $18.57\pm 0.12$ & $0.13\pm 0.02$ \\
2008 August 14.17  & B    & 428  & $18.44\pm 0.08$ & $0.27\pm 0.02$ \\
2008 August 14.17  & V    & 428  & $17.73\pm 0.08$ & $0.30\pm 0.02$ \\
2008 August 15.04  & UVW1 & 727  & $>20.7$         & $<0.02$        \\
2008 August 15.04  & U    & 467  & $18.65\pm 0.13$ & $0.12\pm 0.01$ \\
2008 August 15.04  & B    & 467  & $18.59\pm 0.09$ & $0.24\pm 0.02$ \\
2008 August 15.04  & V    & 467  & $17.65\pm 0.08$ & $0.33\pm 0.02$ \\
2008 August 26.60  & UVW1 & 1522 & $>21.4$         & $<0.01$        \\
2008 August 26.60  & U    & 1024 & $19.99\pm 0.15$ & $0.036\pm 0.005$ \\
2008 August 26.60  & B    & 1024 & $19.16\pm 0.13$ & $0.14\pm 0.01$ \\
2008 August 26.59  & V    & 1024 & $18.60\pm 0.10$ & $0.14\pm 0.01$ \\
2008 October 15.02 & U    & 716  & $>21.2$         & $<0.01$        \\
2008 October 15.02 & B    & 646  & $>21.2$         & $<0.02$        \\
2008 October 15.01 & V    & 716  & $>19.8$         & $<0.04$
\enddata
\tablecomments{Journal of {\it Swift} UV/Optical Telescope
observations of \trans.  Central wavelengths for the various filters
are: 1880 \AA\ (UVW2), 2510 \AA\ (UVW1), 3450 \AA\ (U), 4390 \AA\ (B),
and 5440 \AA\ (V).}
\end{deluxetable}

\clearpage
\begin{deluxetable}{llccc}
\tabletypesize{\footnotesize}
\tablecolumns{5}
\tabcolsep0in\footnotesize
\tablewidth{0pc}
\tablecaption{Optical Photometric Observations \label{tab:opt}}
\tablehead {
\colhead {UT Date}          &
\colhead {Telescope}        &
\colhead {Filter}           &
\colhead {Exposre}          &
\colhead {Magnitude}        \\
\colhead {}                 &
\colhead {}                 &
\colhead {}                 &
\colhead {(s)}              &
\colhead {}                 
}
\startdata
2008 May 31.44  & Magellan & g & 5   & $15.29\pm 0.04$ \\
2008 May 31.40  & Magellan & r & 10  & $14.59\pm 0.04$ \\
2008 May 31.44  & Magellan & i & 5   & $14.28\pm 0.04$ \\
2008 May 31.44  & Magellan & z & 10  & $14.06\pm 0.04$ \\
\enddata
\tablecomments{Journal of optical observations of \trans\ from
May 31.4 UT.}
\end{deluxetable}

\clearpage
\begin{deluxetable}{llccccc}
\tabletypesize{\footnotesize}
\tablecolumns{7}
\tabcolsep0in\footnotesize
\tablewidth{0pc}
\tablecaption{Line Identification in the Low-Resolution Spectra
\label{tab:ldss}}
\tablehead {
\colhead {Line}             &
\colhead {Date}             &
\colhead {Wavelength}       &
\colhead {Velocity}         &
\colhead {EW}               &
\colhead {FWHM}             &
\colhead {Flux}             \\
\colhead {}                 &
\colhead {}                 &
\colhead {(\AA)}            &
\colhead {(km s$^{-1}$)}    &
\colhead {(\AA)}            &    
\colhead {(\AA)}            &    
\colhead {(erg cm$^{-2}$ s$^{-1}$)}                
}
\startdata
\ion{Ca}{2} H          & May 31  & 3938.84 & 310   & $10.5$   & 12.1 & \nod                 \\
                       & June 22 & 3938.30 & 270   & $7.5$    & 8.7  & \nod                 \\ 
\ion{Ca}{2} K          & May 31  & 3973.78 & 315   & $7.9$    & 9.3  & \nod                 \\
                       & June 22 & 3973.47 & 295   & $5.5$    & 7.0  & \nod                 \\ 
H$\delta$              & May 31  & 4102.51 & $-30$ & $-4.7$   & 14.1 & $9.8\times 10^{-15}$ \\
                       & June 22 & 4104.32 & 105   & $-5.7$   & 6.7  & $7.0\times 10^{-15}$ \\ 
H$\gamma$              & June 22 & 4343.25 & 110   & $-5.6$   & 7.2  & $8.6\times 10^{-15}$ \\
H$\beta$               & May 31  & 4864.35 & 100   & $-3.9$   & 9.0  & $1.3\times 10^{-14}$ \\
                       & June 22 & 4864.28 & 100   & $-11.2$  & 6.7  & $2.5\times 10^{-14}$ \\ 
?                      & May 31  & 5857.10 & \nod  & $0.2$    & 7.2  & \nod                 \\
                       & June 22 & 5856.00 & \nod  & $0.6$    & 7.9  & \nod                 \\ 
\ion{He}{1} 5877.25    & May 31  & 5878.96 & 90    & $-1.0$   & 9.8  & $3.9\times 10^{-15}$ \\
                       & June 22 & 5880.13 & 145   & $-2.6$   & 7.6  & $7.3\times 10^{-15}$ \\ 
\ion{Na}{1} D          & May 31  & 5895.65 & 55    & $1.0$    & 10.0 & \nod                 \\
                       & June 22 & 5894.75 & 10    & $2.1$    & 10.2 & \nod                 \\ 
\ion{Fe}{2} 5993.03    & May 31  & 5999.51 & 325   & $-0.5$   & 9.9  & $2.1\times 10^{-15}$ \\ 
                       & June 22 & 5998.22 & 260   & $-0.7$   & 7.1  & $2.0\times 10^{-15}$ \\ 
\ion{Fe}{2} 6434.46    & May 23  & 6437.31 & 130   & $-0.7$   & 21.3 & $3.7\times 10^{-15}$ \\
                       & May 31  & 6434.78 & 15    & $-0.8$   & 20.7 & $3.1\times 10^{-15}$ \\
                       & June 5  & 6436.56 & 100   & $-0.9$   & 18.5 & $3.2\times 10^{-15}$ \\
                       & June 22 & 6438.13 & 170   & $-1.0$   & 9.0  & $3.1\times 10^{-15}$ \\
\ion{Fe}{2} 6517.85    & May 23  & 6523.81 & 270   & $-0.3$   & 7.3  & $1.4\times 10^{-15}$ \\
                       & May 31  & 6522.95 & 235   & $-0.6$   & 9.6  & $2.1\times 10^{-15}$ \\
                       & June 5  & 6522.66 & 220   & $-0.5$   & 7.5  & $1.9\times 10^{-15}$ \\ 
                       & June 22 & 6521.58 & 170   & $-1.4$   & 8.0  & $4.3\times 10^{-15}$ \\
H$\alpha$              & May 23  & 6569.94 & 240   & $-20.2$  & 16.2 & $1.0\times 10^{-13}$ \\
                       & May 31  & 6568.86 & 190   & $-23.2$  & 15.3 & $9.2\times 10^{-14}$ \\
                       & June 5  & 6568.74 & 190   & $-27.4$  & 14.9 & $9.3\times 10^{-14}$ \\
                       & June 22 & 6468.02 & 155   & $-46.0$  & 13.3 & $1.5\times 10^{-13}$ \\
\ion{He}{1} 6679.99    & May 23  & 6684.60 & 210   & $-0.7$   & 17.5 & $3.2\times 10^{-15}$ \\
                       & May 31  & 6684.99 & 225   & $-0.8$   & 13.0 & $3.0\times 10^{-15}$ \\
                       & June 5  & 6684.64 & 210   & $-0.7$   & 8.9  & $2.5\times 10^{-15}$ \\
                       & June 22 & 6684.46 & 200   & $-1.6$   & 11.5 & $5.1\times 10^{-15}$ \\
\ion{He}{1} 7067.14    & May 23  & 7068.33 & 50    & $-0.2$   & 6.4  & $8.3\times 10^{-16}$ \\
                       & May 31  & 7068.51 & 60    & $-0.3$   & 5.4  & $9.4\times 10^{-16}$ \\
                       & June 5  & 7069.89 & 120   & $-0.4$   & 6.6  & $1.4\times 10^{-15}$ \\
                       & June 22 & 7071.10 & 170   & $-0.7$   & 10.7 & $2.1\times 10^{-15}$ \\
$[{\rm CaII}]$ 7293.48 & May 23  & 7299.60 & 250   & $-4.7$   & 4.3  & $2.0\times 10^{-14}$ \\
                       & May 31  & 7299.20 & 235   & $-5.2$   & 5.5  & $1.8\times 10^{-14}$ \\
                       & June 5  & 7299.45 & 245   & $-3.6$   & 4.0  & $1.1\times 10^{-14}$ \\ 
                       & June 22 & 7298.22 & 195   & $-5.4$   & 5.2  & $1.6\times 10^{-14}$ \\
$[{\rm CaII}]$ 7325.91 & May 23  & 7332.04 & 250   & $-4.5$   & 5.4  & $1.9\times 10^{-14}$ \\
                       & May 31  & 7331.62 & 235   & $-4.6$   & 6.5  & $1.6\times 10^{-14}$ \\
                       & June 5  & 7331.80 & 240   & $-3.7$   & 5.2  & $1.1\times 10^{-14}$ \\
                       & June 22 & 7330.61 & 195   & $-4.8$   & 6.1  & $1.4\times 10^{-14}$ \\
\ion{Fe}{2} 7464.44    & May 23  & 7468.98 & 180   & $-0.8$   & 24.5 & $3.4\times 10^{-15}$ \\
                       & May 31  & 7468.14 & 150   & $-0.7$   & 16.0 & $2.4\times 10^{-15}$ \\
                       & June 5  & 7467.85 & 135   & $-1.0$   & 20.1 & $3.0\times 10^{-15}$ \\
                       & June 22 & 7467.54 & 125   & $-1.4$   & 19.2 & $4.0\times 10^{-15}$ \\
\ion{Fe}{2} 7713.83    & May 23  & 7723.98 & 395   & $-1.0$   & 24.2 & $4.0\times 10^{-15}$ \\
                       & May 31  & 7720.87 & 275   & $-0.7$   & 13.8 & $2.5\times 10^{-15}$ \\
                       & June 5  & 7721.19 & 285   & $-0.7$   & 11.6 & $2.1\times 10^{-15}$ \\
                       & June 22 & 7720.93 & 275   & $-1.1$   & 11.1 & $3.1\times 10^{-15}$ \\
\ion{O}{1} 7774.08$^a$ & May 23  & 7780.29 & 240   & $1.4$    & 8.9  & \nod                 \\
                       & May 31  & 7778.85 & 185   & $1.9$    & 10.6 & \nod                 \\
                       & June 5  & 7778.87 & 185   & $1.9$    & 10.3 & \nod                 \\
                       & June 22 & 7776.96 & 110   & $2.0$    & 9.4  & \nod                 \\
\ion{O}{1} 8448.57$^b$ & May 23  & 8454.78 & 220   & $0.5$    & 4.9  & \nod                 \\
                       & May 31  & 8452.95 & 155   & $0.4$    & 5.6  & \nod                 \\
                       & June 5  & 8452.76 & 150   & $0.4$    & 5.8  & \nod                 \\
                       & June 22 & 7776.96 & 110   & $2.0$    & 9.4  & \nod                 
\enddata
\tablecomments{Emission and absorption features identified in the low
resolution spectra of NGC300-2008-OT1.}
\end{deluxetable}

\clearpage

\clearpage

\clearpage

\clearpage

\clearpage

\clearpage

\clearpage

\clearpage

\clearpage

\clearpage

\clearpage

\clearpage

\clearpage

\clearpage

\clearpage

\clearpage

\clearpage

\clearpage

\clearpage

\clearpage

\clearpage

\clearpage

\clearpage

\clearpage

\clearpage

\clearpage

\end{document}